\documentclass[twocolumn]{aastex631}

\usepackage[varvw]{newtxmath}

\usepackage[T1]{fontenc}
\usepackage{ae,aecompl}
\usepackage[normalem]{ulem} 
\usepackage{comment}

\usepackage{amsmath}	
\usepackage{amssymb}	
\usepackage{bm}
\usepackage{graphicx}	
\usepackage[caption=false]{subfig}
\usepackage{xcolor} 

\newcommand{\BB}{{\bm B}}
\newcommand{\beq} {\begin{equation}}
\newcommand{\cs}{c_{\rm s}}
\newcommand{\eeq} {\end{equation}}
\newcommand{\eg}{e_{\rm g}}
\newcommand{\er}{e_{\rm r}}
\newcommand{\kms}{{\rm km~s}^{-1}}
\newcommand{\jR}{$j$-$R$}
\newcommand{\nth}{n_{\rm th}}
\newcommand{\pcc}{{\rm cm}^{-3}}
\newcommand{\rr}{{\bm r}}
\newcommand{\tdef}{t_{\rm def}}
\newcommand{\uu}{{\bm u}}
\newcommand{\vrot} {v_{\rm rot}}

\definecolor{myblue}{RGB}{0, 90, 189}

\shorttitle{AASTeX v6.3.1 Sample article}
\shortauthors{Arroyo-Ch\'avez and V\'azquez-Semadeni}

\begin{document}
\title[Angular momentum evolution during fragmentation]{Evolution of the angular momentum during gravitational fragmentation of molecular clouds\footnote{Released on June, 1st, 2021}}


\author[0000-0002-7082-0587]{Griselda Arroyo-Ch\'avez}
\author[0000-0002-1424-3543]{Enrique V\'azquez-Semadeni}
\affiliation{Instituto de Radioastronom\'ia y Astrof\'isica \\
PO Box 3-72. 58090 \\
Morelia, Michoac\'an, M\'exico}






\begin{abstract}
We investigate the origin of the observed scaling $j\sim R^{3/2}$ between the specific angular momentum $j$ and the radius $R$ of molecular clouds (MCs) and their their substructures, and of the observed near independence of $\beta$, the ratio of rotational to gravitational energy, from $R$. To this end, we measure the angular momentum (AM) of sets of particles in an SPH simulation of the formation, collapse and fragmentation of giant MCs. The sets of SPH particles are defined either as ``clumps'' (connected particle sets), or as lagrangian sets that conform a connected clump only at a certain time $\tdef$. We find that: {\it i)} Clumps evolve along the observed \jR\ relation at all times, {\it ii)} Lagrangian particle sets evolve along the observed relation when the volume containing them also contains a large number of other ``intruder'' particles. Otherwise, they evolve with $j\sim$ cst. {\it iii)} Tracking lagrangian sets to the future, we find that a subset of the SPH particles participates in the collapse, while another disperses away. {\it iv)} Noting that, under AM conservation, $\beta$ increases during contraction, we suggest that its near independence of radius may arise from the competition between this increase and an increase in the AM exchange rate at higher rotational energy density gradient. {\it v)} We suggest that, if MCs are globally dominated by gravity, the observed \jR\ relation arises because the observational selection of its dense structures amounts to selecting the fragments that have lost AM via Reynolds stresses from their neighbors.




\end{abstract}




\section{Introduction}
\label{sec:introduction}


Large-scale velocity gradients have been observed in clouds and their clums and cores since the 1970s, and have usually been interpreted as evidence of rotation of the clouds \citep[e.g.,] [] {Belloche2013}. In particular, \citet{Fleck.Clark1981} found that the angular velocity, $\Omega$, has a dependence on the radius of the cloud of the form $\Omega (R) \propto R^{p}$, with $p \sim -2/3$,. These authors attributed this result to the turbulent cascade generated from the galactic differential rotation, driven by the shearing motions in the galacic disk, while \citet{Goldsmith.Arquilla85} found that for their sample, the specific angular momentum (hereafter SAM), $j = J / M$, where $J$ is the total AM and $M$ is the cloud's mass, scales with the radius as $R^{1.4}$, so that $\Omega \propto R^{-0.6}$, in agreement with the results of \citet{Fleck.Clark1981}. The interpretation by \citet{Goldsmith.Arquilla85} was that this relation is evidence of the loss of AM during the contraction and fragmentation of the clouds, suggesting that this loss is due to the redistribution of AM in the orbital motions of the fragments. 

These works illustrate what is known as the ``angular momentum problem'' \citep[e.g.,] [] {Spitzer78, Bodenheimer95}, which consists in the apparent loss of SAM form molecular cloud scales ($\sim 10$ pc) to the scales of dense cores ($\sim 0.1$ pc) and protostellar disks ($\lesssim 0.01$ pc), as illustrated in Fig.\ \ref{fig:data}. Therefore, the angular momentum problem is actually the problem of how AM is redistributed as a cloud contracts gravitationally and fragments.



At the scales of giant molecular clouds, \citet{Imara.Blitz2011} have shown that the position angles of the velocity gradients of molecular gas and of the diffuse gas are highly divergent from each other, and interpreted as rotation, have magnitudes implying that the SAM of the cloud is less than that of the surrounding medium. At the scale of dense cores, the decreasing nature of the SAM implied by the velocity gradients has been known since the work of \citet[] [hereafter G93] {Goodman+93}, with a scaling $j \sim R^{1.6}$. Moreover, assuming that {\it i)} the ratio $\beta$ of rotational to gravitational energy was constant for all cores, {\it ii)} a linewidth-size scaling relation of the form $\sigma \propto R^{1/2}$ \citep{Larson81} holds for the cores, and {\it iii)} the cores are in approximate virial equilibrium, G93 were able to analytically derive a scaling relation of the form $j \sim R^{3/2}$.


On the even smaller scales of protoplanetary discs, \citet{Pineda+2019} have found that, for three sources in Perseus containing young stellar objects, the radial profile of SAM on scales $\sim 10^{3}$--$10^{4}$ AU scales as $R^{1.8 \pm 0.04}$. More recently, using a sample of 11 objects from the {\it IRAM CALYPSO} catalog of dense gas observations, \citet{Gaudel+2020} have found that the radial profile of SAM on scales from $50$ to $5000$ AU exhibits two regimes inside the protostellar envelope: one in which the SAM scales with radius as $j \propto R^{1.6 \pm 0.2}$ on scales $\gtrsim 10^3$ AU, and another in which $j$ tends to be constant, on scales $\sim 50$--$10^3$ AU.

The angular momentum problem has received various tentative solutions within the context of different models for molecular clouds and star formation (SF). In the model of magnetically-supported clouds, with star formation mediated by ambipolar diffusion \citep[see, e.g., the reviews by] [] {Shu+87, Mouschovias91}, the preferred mechanism to explain the redistribution of AM was magnetic braking \citep[see the review by] [and references therein] {Bodenheimer95}. This mechanism consists of the torsion applied, by the rotation of the structure, to the magnetic field lines that permeate both the structure and the surrounding medium, therefore transporting the AM of the structure outwards through the tension of the field lines. However, at present it is known that the magnetic support theory presents a series of problems, and has been superseded by the so called ``gravoturbulent'' scenario,
in which clouds are supported against their self-gravity by the pressure exerted by their internal supersonic turbulence \citep[e.g.,] [] {MacLow.Klessen2004}. The compressions, however, produce density enhancements that may locally exceed their own Jeans masses, and proceed to collapse \citep[e.g.,] [] {VS+03, MacLow.Klessen2004}

Within the context of an inhomogeneous, self-gravitating, rotating cloud, \citet{Larson84} proposed that gravitational torques exerted by non-radial gravitational forces originating from sheared density fluctuations could be responsible for the AM transfer. More recently, \citet{Li_P+2004}, have shown that the cores formed in Super-Alfv\'enic turbulent simulations follow the relation $j \sim R^{3/2}$. However, \citet{Jappsen.Klessen04} showed that non-magnetic, continuosly-driven turbulent numerical simulations also exhibit a $j \sim R^{3/2}$ scaling, supporting the view that this relation, and the associated AM transfer mechanism, are not due to the magnetic field. Like \citet{Larson84}, these authors also invoked gravitational torques as the mechanism responsible for AM transfer, although no tests were performed to support this claim.


On the other hand, \citet {VS+19} have recently discussed a number of problems of the gravoturbulent scenario. Chief among them is the inconsistency \citep{IM+16} between the velocity dispersion-size scaling expected for turbulence ($\sigma \propto R^{1/2}$) and the observed scaling, $\sigma \propto (\Sigma R)^{1/2}$, which involves in addition a dependence of the velocity dispersion on the column density, $\Sigma$. The latter scaling is expected for either virial equilibrium \citep{Heyer+09} or collapse \citep{BP+11}. However, because this scaling is observed across all scales from giant molecular clouds to massive dense cores \citep[e.g.,] [] {Heyer+09, BP+11, Kauffmann+13, Leroy+15, Miville+17, Traficante+18, BP+18}, it does not appear feasible that {\it all} size and density scales are virialized while simultaneously producing (non-virialized) density fluctuations in their interiors. Therefore, it has been suggested that molecular clouds and their substructures are all in a state of non-homologous gravitational contraction and fragmentation \citep[e.g.,] [] {Hartmann_Burkert07, VS+09, VS+19, BP+11, IM+16}, in a regime of global, hierarchical collapse \citep[GHC;] [] {VS+19}. However, the origin of the observed $j$-$R$ scaling and the AM transfer mechanism have not been investigated within the context of the GHC scenario and numerical simulations.
This is the objective of the present work, in which we follow the evolution of the AM in dense clumps defined and tracked over time in different ways in a numerical simulation of GHC, in order to understand the mechanism of AM exchange and determine the degree of consistency with the observed scaling.

The structure of the paper is as follows. In Sec.\ \ref{sec:AM_transf_mech} we first revisit the sources of torques acting on a finite-size fluid parcel with respect to a given origin, emphasizing the role of turbulent eddy viscosity. Next, in Sec.\ \ref{sec:observational data}, we collect data from several observational studies concerning the $j$-$R$ scaling for structures over two orders of magnitude in size, as a guideline for the expected scaling. In Sec.\ \ref{Sec:Numerical Data} we describe the main features of the numerical simulation used in this work and the prescriptions for defining and time-tracking the clumps. In Sec.\ \ref{sec:results} we present our measurements, and in Sec.\ \ref{sec:discussion} we discuss the implications of our results, revisit G93's derivation of the \jR\ scaling within the context of GHC, and discuss the possible origin of the near independence of $\beta $ with $R$. Finally, in section Sec.\ \ref{sec:conclusions}, we present our main conclusions.


\section{On the nature of the AM transfer mechanism: the available torques} \label{sec:AM_transf_mech}

In order to put the remainder of the paper in context, we start by writing the equation governing the evolution of the AM of a fluid parcel of volume $V$ with respect to some coordinate origin. This is formally done by taking the cross product of the momentum conservation equation with the position vector $\rr$ and integrating over $V$:
\begin{align}
 \int_{V} {\bm r} \times  \frac{\partial {(\rho \uu)}}{\partial t} dV = - \int_{V}  {\bm r} \times \nabla \cdot (\rho \uu\uu) dV - \int_{V} {\bm r} \times  \nabla P dV \nonumber & \\ - \int_{V} {\bm r} \times  \rho \nabla \phi dV + \int_{V} {\bm r} \times  \mu  (\nabla^{2} \uu + \nabla \nabla \cdot \uu\uu) dV \nonumber & \\ + \int_{V} {\bm r} \times \frac{1}{4 \pi} (\nabla \times \BB) \times \BB dV, 
\label{eq:total torque}
\end{align}
In this equation, the terms on the right-hand side are the torques acting on the fluid parcel, such as gravitational torques (third term), viscous torques (fourth term), torques by pressure gradient (second term), magnetic torques (last term), and what we shall call ``ram-pressure'' or ``hydrodynamic'' torques, given by the first term. This term originates from the advection term of the momentum equation, which in general represents the transfer of momentum in the $i$ direction by the velocity in the $j$ direction, where $i$ and $j$ are any two of the coordinate axes. In eq.\ (\ref{eq:total torque}), this translates to the torque exerted by these momentum exchanges when referred to some coordinate origin. 

When an averaging of the momentum equation is performed to separate the mean flow from the turbulent fluctuations, the nonlinear term gives rise to appearance of the so-called Reynolds stress term, which is responsible for the loss of {\it linear} momentum from the mean motion \citep[see, e.g.,] [Ch.\ 4] {Lesieur08}. In principle, one can take the cross product with the averaged equations and then compute the torques due to the Reynolds stresses. So, the hydrodynamic torque term in eq.\ \eqref{eq:total torque} is related to the torques exerted by the Reynolds stresses.

Another noteworthy feature of eq. (\ref{eq:total torque}), is that it can be seen as an intermediate, vector expression between the scalar and tensor forms of the virial theorem (VT), since all of these involve products of the position vector with the momentum equation, integrated over volume. The scalar VT involves the dot product, to obtain the work done by the forces, while the tensor VT involves the direct product (dyadic) between $\rr$ and the momentum equation. Equation (\ref{eq:total torque}) involves the cross product, giving a vector, which is the net torque on the fluid parcel $V$, although it also has dimensions of energy.


The role of hydrodynamic and pressure gradient torques seems to have been somewhat neglected in the framework of molecular clouds, clumps and cores. In the study of accretion disks, the AM transfer mechanism constitutes the fundamental process underlying the ability of the disk to transfer mass to the star. Viscous torques are known to be way too small to be important \citep[e.g.,] [] {Hartmann09}, and therefore turbulent, Reynolds stress torques are generally invoked \citep{SS73}. Since the actual turbulent velocity dispersion and eddy size scale are unknown in the disks, the effect of these torques is modeled via an eddy viscosity $\nu_v$, which is a coefficient relating the Reynolds stresses to the mean flow, and is given by
\begin{equation}
\nu_v = \alpha \cs H,
\label{eq:alpha_model}
\end{equation}
where $\cs$ is the sound speed, $H$ is the scale height of the disk \citep{SS73}, and $\alpha$ is an adjustable parameter. However, since keplerian disks are Rayleigh stable, the rotational instability cannot be expected to drive the turbulence, and alternative driving mechanisms must be sought, most noticeable among them being the magneto-rotational instability \citep{Balbus_Hawley91}.


On the other hand, in molecular clouds and their clumps and cores, the AM transfer mechanism is not clear, since, similarly to the case of disks, the molecular viscosity is negligible. In a seminal paper, \citet{Larson84} argued against eddy viscosity on the basis that no mechanism capable of sustaining the turbulence in a cloud or clump was known at the time, and argued in favour of gravitational torques instead. These were assumed to arise from the gravitational force of the density fluctuations within the cloud or clump.  

However, at present there is a consensus that the clouds, clumps and cores are in general turbulent, so there is no shortage of turbulence in these objects, even if the precise mechanism responsible for driving it is still a matter of debate. One likely source of turbulence is the very gravitational contraction of these objects \citep[e.g.,] [] {Vazquez-Semadeni+98, Klessen_Hennebelle10, Robertson_Goldreich12, Murray_Chang15, Xu_Lazarian20, Guerrero.Vazquez2020}, which may operate from the scale of dense cores and downwards \citep[within the context of the gravoturbulent scenario of the clouds;] [] {MacLow.Klessen2004}, or even starting at the scale of whole GMCs \citep[within the context of the global hierarchical collapse scenario, GHC;] [] {VS+19}. \citet{Guerrero.Vazquez2020} have estimated through numerical simulations that turbulence driven by gravitational contraction may contain roughly half the kinetic energy as the infall motions.

In the remainder of this paper, we will investigate the transfer of AM among fluid parcels in the cloud in SPH numerical simulations and, although we do not explicitly demonstrate that the mechanism is ram and thermal pressure gradient torques, there appears to be no {\it a priori} reason to rule them out, either. In a future contribution we intend to compare simulations with and without self-gravity, in order to determine the role of each type of torque acting on the fluid.



\section{The observed \jR\ scaling}
\label{sec:observational data}

\begin{figure}
 \includegraphics[width=\columnwidth]{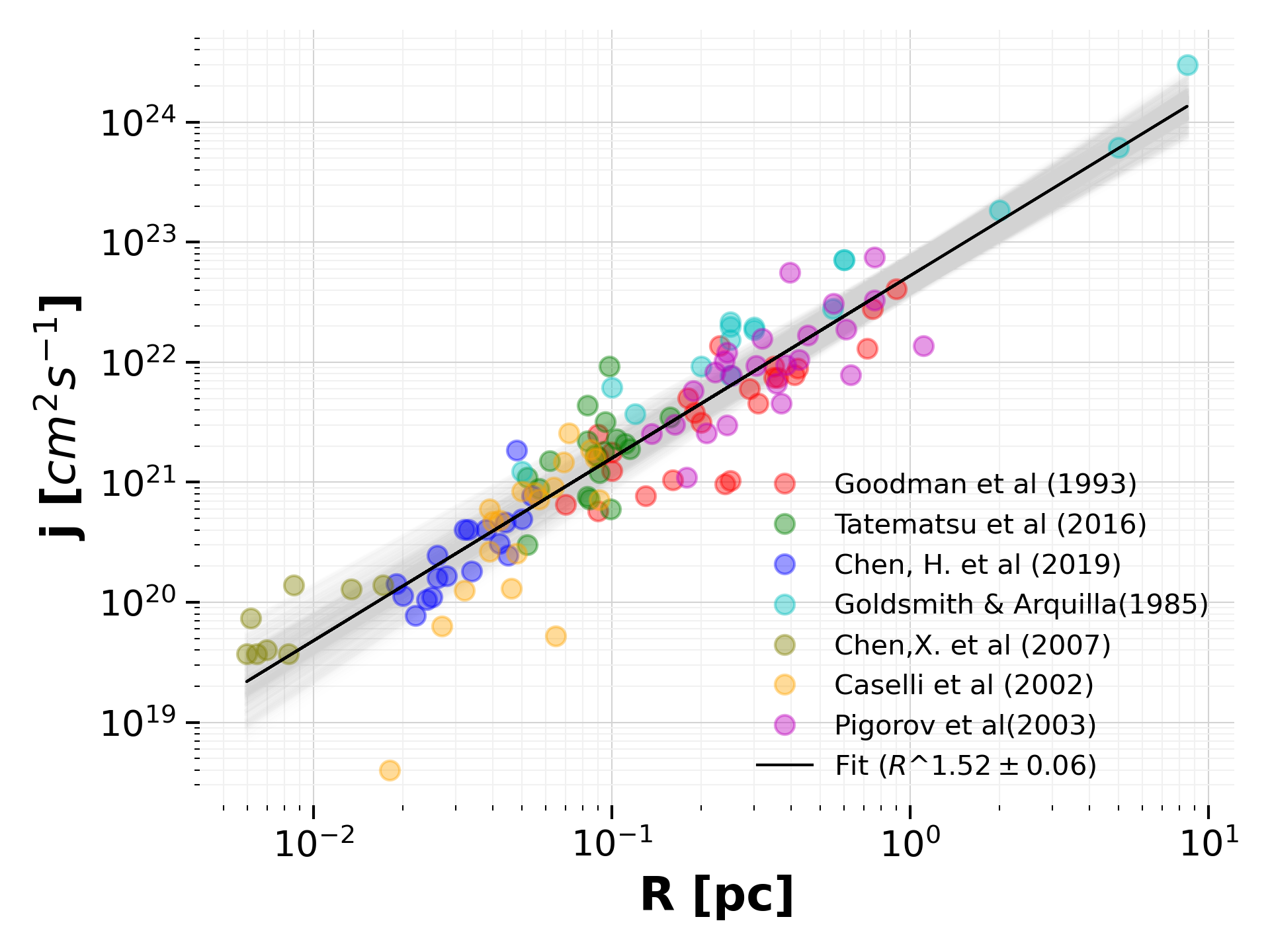}
 \caption{The observed \jR\ relation for molecular structures of sizes $\sim 0.01$--10 pc.
The solid line represents the best fit. The set of light-gray lines represent the 1-$\sigma$ error.
}
 \label{fig:data}
\end{figure}

In order to obtain a statistically significant observational guideline for the $j$-$R$ relation we will obtain from our numerical simulation, in Fig. \ref{fig:data} we have compiled the measurements of $ j $ presented by \citet{Goldsmith.Arquilla85}, \citet{Goodman+93}, \citet{Chen_Hope+2019b}, \citet{Chen_X+2007}, \citet{Caselli+2002}, \citet{Pigorov+2003}, and \citet{Tatematsu+16}, for clouds and clumps of sizes ranging from $\lesssim 0.01$ pc to $\gtrsim 10$ pc. 
A least squares fit to the data in this figure gives the expression
\begin{equation}
    j = 10^{22.7 \pm 0.002} \left(\frac{R}{1 {\rm pc}}\right)^{1.52 \pm 0.06} {\rm cm}^2~ {\rm s}^{-1},
    \label{eq:numerical fit}
\end{equation}
which is represented by the black line in the figure.
The associated $1\sigma$ error in the fitting parameters is indicated by the shaded region. 

It should be noted that these data were obtained from the measurement of velocity centroid gradients in the clouds, which are usually interpreted as  rotation \citep{Kutner+77, Phillips99, Rosolowsky+03}, although they can just as well be interpreted as due to expansion, contraction or shear, or combinations thereof \citep{Belloche2013,Tobin_J+2012a}. In particular, it is important to note that at least part of this gradient {\it must} correspond to converging motions, as convergence of the velocity field is required by the continuity equation in order to produce the density enhancement constituting a clump. Therefore, in what follows, we shall assume that some undetermined but non-negligible fraction of the observed large-scale velocity gradient in the clouds corresponds to rotation, while the rest of the kinetic energy may correspond to convergence or to turbulent motions.


It is important to note that the methods and molecular tracers used to obtain the values of the radius and SAM may differ between the observational samples collected in this section. Thus, for example, while \citet{Goodman+93} obtained the velocity gradient by fitting a solid-body rotation to the observed $v_{\rm LSR}$ map of a cloud, \citet{Tatematsu+16} measured the velocity gradient using position–velocity diagrams passing through core centers, and made sinusoidal fits against the position angle. This gives rise to the possibility that the same clump has two different values of $j$ and $R$ in two different samples, which implies that the same object may appear more than once in Fig.\ \ref{fig:data}. 



\section[]{Numerical data}
\label{Sec:Numerical Data}

In this section we describe the numerical simulation as well as the procedures we used to follow the evolution of the SAM for the selected clumps, as well as the way in which we determine the physical properties of the clumps.


\subsection{The simulation}
\label{subsec: Simulation}

We use a simulation of decaying turbulence in the warm neutral atomic gas first presented in \citet{Heiner+15}. The simulation was performed using \textsc{Gadget-2} \citep{Springel+01}, a smoothed-particle hydrodynamics (SPH) code, using $296^{3} \approx 2.6 \times 10^{7}$ particles in a box of $256$ pc per side. With a constant mass per particle set at $0.6 M_{\odot}$, the total mass in the box is $1.58 \times 10^{6} M_{\odot}$. The initial density and temperature were set at $n(t=0) = 3\, \pcc$ and $T(t=0) = 730$ K, respectively. This initial density is intended to represent the density typical in a Galactic spiral arm, and the initial temperature corresponds to thermal equilibrium between heating and cooling at this density.  

In this version of the code, the prescription for sink particles from \citet{Jappsen+05} was used, setting the density threshold for sink particle formation at $3.2 \times 10^{6}\, \pcc$. Also included were the cooling and heating functions from \citet{KI02}, with the typographical correction given by \citet{Vazquez-Semadeni+07}, as well as the rpSPH algorithm from \citet{Abel11}, which reports improvements in the management of physical instabilities such as Kelvin-Helmholz and Rayleigh-Taylor, eliminating some non-physical effects present in the standard SPH prescription. No prescription for stellar feedback is included, so we only considered clumps with a low enough sink particle content that the omission of feedback does not render them unrealistic (cf.\ Sec.\ \ref{subsec: Def and search}).

The simulation is initially driven with purely solenoidal modes in the wavenumber range $1 \le k \le 4$ until $t=0.65$ Myr, and then left to decay. At that time, a maximum velocity dispersion of $\sigma \approx 18~ \kms$ is reached. Although this velocity dispersion is somewhat high compared to actual values \citep[$\sim 8$--$10~ \kms$; e.g.,] [] {HT03}, this is partially compensated by the purely solenoidal character of the fluctuations, which induces less compressions than a mixture of solenoidal and compressible modes \citep[e.g.,] [] {VS+96, Federrath+08}. In addition, by the time steps in which the simulation is analysed, the velocity dispersion has dropped to $ \sim 4~ \kms$.


\subsection{Clump definition}
\label{subsec: Def and search}

In the context of this paper, we will use the term ``clump'' in a generic way to denote any overdensity above a specified threshold ($n_{\text{th}}$). Thus, clumps defined at the highest thresholds (and generally more compact) will correspond to ``cores'', while clumps defined at the lowest thresholds (and usually more extended) will correspond to ``clouds''. The clumps are initially defined using the algorithm introduced in \citet{Camacho+16}, as a ``connected'' set of SPH particles (i.e., within their smoothing lengths) above some specified density threshold, around a local maximum of density. Clumps defined in this way can contain substructures; i.e., a single ``cloud'' may contain several ``cores''. 

Our procedure is similar to the {\sc Dendrograms} algorithm, except that we do not produce a structure tree with the ``lineage'' of the clumps, and the thresholds are arbitrary, rather than being set exactly at the level where a large structure fragments into smaller ones. \citet{Camacho+20} have shown that varying the threshold levels basically changes the types of object selected, but does not significantly affect the general trend of the ensemble of objects.


\subsection{The numerical clump sample}
\label{subsec:sample}

We select four timesteps in the simulation after the formation of the first sink ($t=14.74$ Myr) to analyze the $j$--$R$ relation of a sample of clumps over the entire numerical box at each time. The timesteps correspond to $t=17.92$, $t=19.92$, $t=23.24$ and $t=25.23$ Myr.
We apply the clump-finding algorithm at these times with thresholds $n_{\text{th}} = 10^{3}$, $3 \times 10^{3}$, $10^{4}$, $3 \times 10^{4}$, and $10^{5}$ cm${}^{-3}$, rejecting those clumps with less than 60 particles (1.5 times the number of particles within the smoothing radius) in order to guarantee that the retained ones are sufficiently resolved \citep{Bate_Burkert97}. This implies a mass of at least $3.6$ $M_{\odot}$ per clump. Furthermore, we only keep the clumps with column density $\Sigma \gtrsim 30$ M${}_{\odot}$pc${}^{-2}$, since both observations \citep[e.g.,] [] {Keto_Myers86, Leroy+15, Traficante+18} and simulations \citep[e.g.,] [] {Camacho+16, Mejia-Ibanes+2016} suggest that clumps with lower column densities are mostly dominated by turbulence, while above $\Sigma \gtrsim 30$ M${}_{\odot}$pc${}^{-2}$ gravity is probably dominant.


\begin{figure}
 \includegraphics[width=\columnwidth]{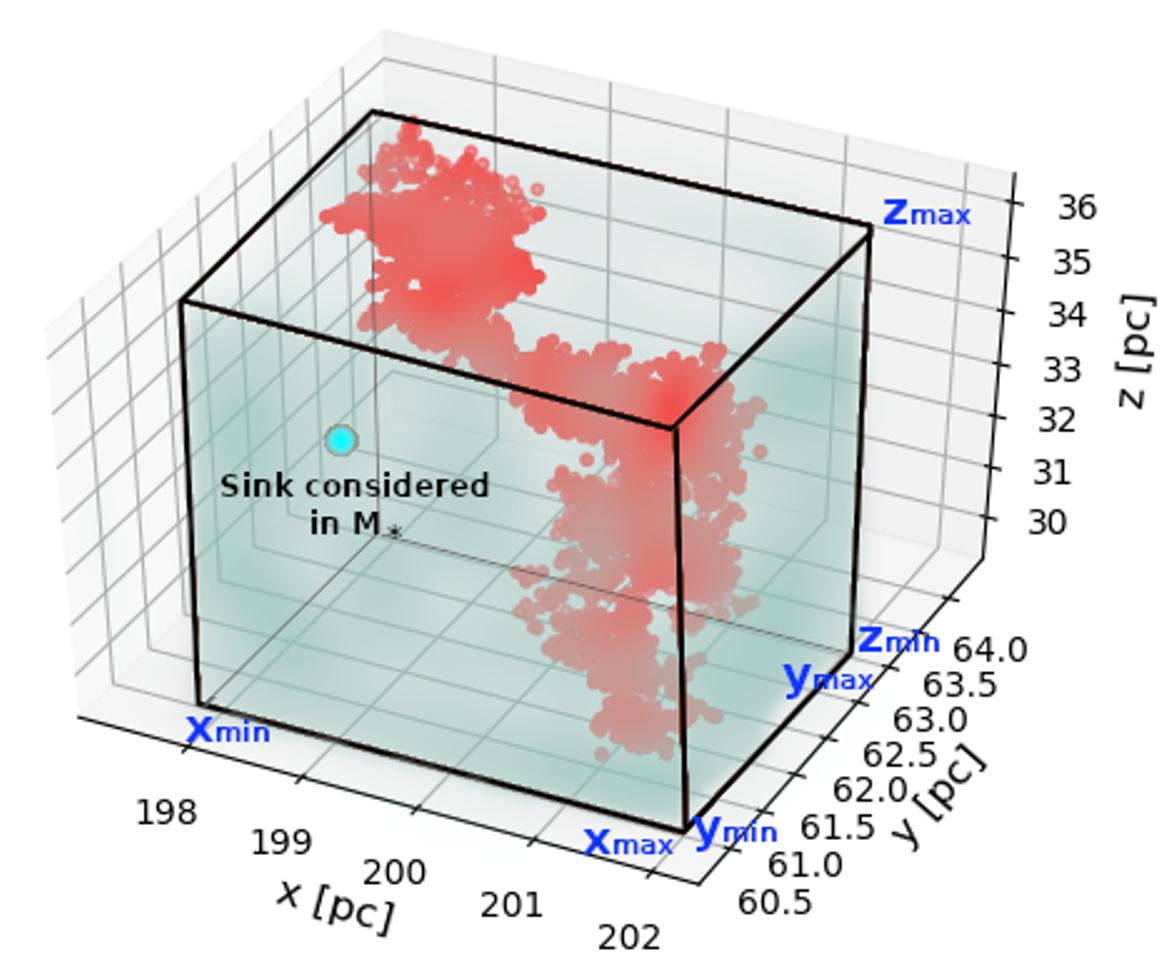}
 \caption{Parallelepiped defined by the minimum and maximum coordinates along each axis of the SPH particles belonging to the clump (red dots). Any sinks (e.g., the cyan point) within this parallelepiped will be considered as a contribution to $M_*$ in eq.\ (\ref{eq:efficiency}).}
 \label{fig:box}
\end{figure}

Since the simulation does not include any form of feedback, it is necessary to establish a criterion to avoid including into account clumps in which the stellar content would be expected to alter their dynamics significantly. We thus restrict our sample to clumps whose star formation efficiency (SFE) satisfies
\begin{equation}
\centering
\frac{M_{*}}{M_{\text{tot}}} \equiv \frac{M_{*}}{M_{*}+M_{\text{gas}}} < 30 \%,
\label{eq:efficiency}
\end{equation}
where $M_{*}$ is the mass in stars (sinks). A sink will be associated with a clump if it is within the box defined by the minimum and maximum values of the positions of the particles on the three coordinate axes, and it will be considered in the calculation of the SFE as a contribution to $M_{*}$ (as can be seen in Fig. \ref{fig:box}).

\subsection{Time-tracking of lagrangian particle sets and of regular clumps (overdensities)} \label{sec:time_tracking}

In order to study the evolution of the SAM of the clumps and their constituent SPH particles, we use two different approaches. In the first, we consider a few randomly-chosen clumps (originally defined as connected overdensities in the flow) from the full sample, and follow the (fixed) set of their constituent particles over time.  We refer to these as {\it lagrangian} SPH particle sets. The time tracking was performed over a few megayears both towards the past and towards the future of the time $\tdef$ at which the sets were defined.
The tracking from the past was carried out over $2.65$ Myr. The tracking to the future was carried out only until the last snapshot before a sink formed within the particle set. This was done because, once a sink forms, it begins to accrete both mass (SPH particles) and AM from the lagrangian set, and thus the variation of the AM ceases to be caused exclusively by exchanges with the set's neighbouring gas parcels. It should be noted that, for a threshold density of $n_{\text{th}} = 10^{3}$ cm${}^{-3}$, the smallest  threshold used to define clumps in this work, the free fall time corresponds to $\sim 1.44$ Myr. Thus, the tracking time intervals we use guarantee that all clumps have ample time to evolve dynamically.

The second approach is the traditional one, in which we define the clumps as connected overdensities at all times during the tracking. It is very important to note that, with this definition, {\it the clumps do not consist of the same particles at the various times}. In fact, the clumps, defined over the same density threshold $\nth$ at the various times, tend to increase their mass. We considered three clumps at times $t=17.26$ and $17.93$ Myr. The characteristics of all the clumps tracked in time in this work are compiled in Table \ref{tab:all clumps}.

\begin{table*} 

\caption{Characteristics of clumps tracked over time}
\hspace{-1.1cm}
\begin{tabular}{|c|c|c|c|c|c|c|}
\hline
\multicolumn{3}{|c|}{}                                                  & \multicolumn{2}{c|}{Type}                                                                                                                      & \multicolumn{2}{c|}{Tracking}                                                                                                                                                                                          \\ \hline
Name & $t_{\rm def}$ (Myr) & $n_{\rm th}$ (cm${}^{-3}$)                & \hspace{-1cm} \begin{tabular}[c]{@{}c@{}}Lagrangian \\ set\end{tabular} & \hspace{-1cm} \begin{tabular}[c]{@{}c@{}}Conected \\ clump \\ above \\ $n_{\rm th}$\end{tabular} & \hspace{-1cm} \begin{tabular}[c]{@{}c@{}}From \\ past ($t<t_{\rm def}$) \\ to \\ present ($t=t_{\rm def}$)\end{tabular} & \hspace{-1cm} \begin{tabular}[c]{@{}c@{}}From \\ present ($t=t_{\rm def}$)\\ to \\ future ($t>t_{\rm def}$)\end{tabular} \\ \hline
C1   & $19.92$              & $10^{3}$          & X                                                         &                                                                                    & X                                                                                                         &                                                                                                            \\ \hline
C2   & $19.92$              & $3 \times 10^{3}$ & X                                                         &                                                                                    & X                                                                                                         &                                                                                                            \\ \hline
C3   & $19.92$              & $10^{4}$         & X                                                         &                                                                                    & X                                                                                                         &                                                                                                            \\ \hline
C4   & $19.92$              & $ 3 \times 10^{4}$ & X                                                         &                                                                                    & X                                                                                                         &                                                                                                            \\ \hline
C5   & $19.92$              & $10^{5}$          & X                                                         &                                                                                    & X                                                                                                         &                                                                                                            \\ \hline
C6   & $25.23$              & $10^{3}$          & X                                                         &                                                                                    & X                                                                                                         &                                                                                                            \\ \hline
C7   & $25.23$              & $ 3 \times 10^{3}$  & X                                                         &                                                                                    & X                                                                                                         &                                                                                                            \\ \hline
C8   & $25.23$              & $10^{4}$         & X                                                         &                                                                                    & X                                                                                                         &                                                                                                            \\ \hline
C9   & $25.23$              & $3 \times 10^{4}$  & X                                                         &                                                                                    & X                                                                                                         &                                                                                                            \\ \hline
C10  & $25.23$              & $10^{5}$         & X                                                         &                                                                                    & X                                                                                                         &                                                                                                            \\ \hline
C11  & $17.26$              & $10^{3}$          & X                                                         &                                                                                    &                                                                                                           & X                                                                                                          \\ \hline
C12  & $19.92$              & $3 \times 10^{3}$ & X                                                         &                                                                                    &                                                                                                           & X                                                                                                          \\ \hline
C13  & $19.92$              & $10^{4}$          & X                                                         &                                                                                    &                                                                                                           & X                                                                                                          \\ \hline
C14  & $23.24$              & $3 \times 10^{4}$ & X                                                         &                                                                                    &                                                                                                           & X                                                                                                          \\ \hline
C15  & $23.24$              & $10^{3}$          & X                                                         &                                                                                    &                                                                                                           & X                                                                                                          \\ \hline
C16  & $17.26$              & $3 \times 10^{3}$&                                                           & X                                                                                  &                                                                                                           & X                                                                                                          \\ \hline
C17  & $17.93$              & $3 \times 10^{3}$ &                                                           & X                                                                                  &                                                                                                           & X                                                                                                          \\ \hline
C18  & $17.93$              & $3 \times 10^{3}$ &                                                           & X                                                                                  &                                                                                                           & X                                                                                                          \\ \hline
\multicolumn{7}{l}{$t_{\rm def}$: definition time; $n_{\rm th}$: definition threshold density} 
\end{tabular}
\label{tab:all clumps}
\end{table*}

\begin{figure}
 \includegraphics[width=\columnwidth]{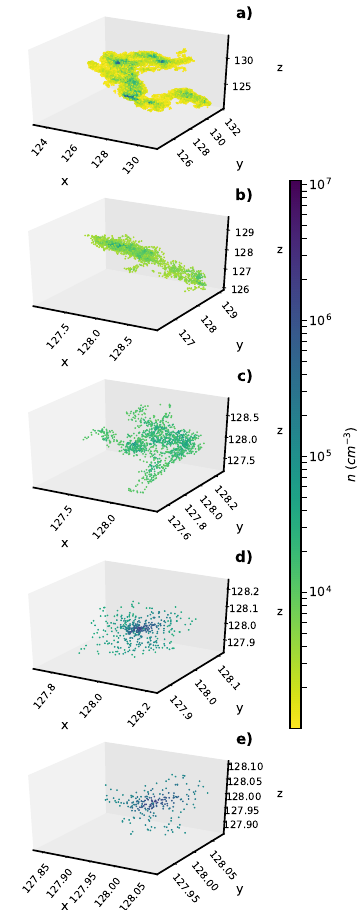}
 \caption{Sequence of hierarchically-nested clumps defined at time $t=19.92$ Myr (clumps C1-C5 in Table \ref{tab:all clumps}), defined at various density thresholds: (a) $n_{\text{th}} = 10^{3} n_{0}$, (b) $n_{\text{th}} = 3 \times 10^{3} n_{0}$, (c) $n_{\text{th}} = 10^{4} n_{0}$, (d) $n_{\text{th}} = 3 \times 10^{4} n_{0}$ and (e) $n_{\text{th}} = 10^{5}  n_{0}$. All clumps are sub-structures of the same cloud. The axes represent the size in pc.}
 \label{fig:muestra}
\end{figure}

As an illustration of the appearance and mass distribution of the clumps that we will be studying in this work, in Fig.\ \ref{fig:muestra} we show a sample of hierarchically-nested clumps (clumps C1-C5 in Table \ref{tab:all clumps}) that were defined through different density thresholds from $\nth = 10^3 \pcc$ to  $\nth = 10^5 \pcc$ at time $t =  19.92 $ Myr in the simulation. It can be seen that the clumps span sizes from over 10 to a few tenths of parsec across this density range.

\subsection{Size estimation}
\label{subsec:radius}

As can be seen in Fig. \ref{fig:muestra}, clumps have an amorphous structure, so specifying a radius for them can be quite an ambiguous task. As a first approximation, we will calculate the radius of a clump as that of a sphere with the same volume; i.e., $R =(3V/4 \pi)^{1/3}$. In turn, for a discrete set of SPH particles, the total volume can be calculated as \citep{Camacho+16}
\begin{equation}
V = \sum_{i = 1}^{N_{c}} V_{i} = \sum_{i = 1}^{N_{c}} \frac{m_{p}}{\rho_{i}}=m_{p} \sum_{i = 1}^{N_{c}} \rho_{i}^{-1}.
\label{eq:total vol}
\end{equation}

It should be noted that this way of computing the radius only applies at the time when all the member particles of the clump are ``connected''; that is, when all the particles are within the smoothing radius of other particles in the set. However, when following a set of particles over time (either to the past or the future), some particles of the set may ``disconnect'', ceasing to be withing the smoothing radius of any other particle of the set. In this case, particles that were not part of the set initially will be now located in-between the original member particles. We will refer to these non/member particles as ``intruders''. Note that the intruders may well be part of a new clump defined by means of a density threshold at the new time of observation, but they were not part of the originally defined clump.


Once intruder particles have penetrated among the original set of member particles, the radius calculated from eq. (\ref{eq:total vol}) will not reflect the true extent of the  volume containing the original member particles. In this case, we will determine the radius as the geometric average of half the maximum difference in position of the constituent particles of the clump along each of the coordinate axes. This is, as

\begin{equation}
R \approx {\left( \frac{x_{max}-x_{min}}{2} \times \frac{y_{max}-y_{min}}{2} \times \frac{z_{max}-z_{min}}{2}\right)}^{1/3}.
\label{eq:geometric average}
\end{equation}


\subsection{Calculation of the specific angular momentum}
\label{subsec:j}

The AM of each clump is calculated from the position and velocity vectors of the SPH particles making up the clump, with respect to its own center of mass, so that,
\begin{equation}
\mathbf{J} = \mathbf{r_{\text{CM}}} \times \mathbf{p_{\text{CM}}} = \mathbf{r_{\text{CM}}} \times m_{\text{p}} \mathbf{v_{\text{CM}}},
\label{eq:ang_mom}
\end{equation}
where the subscript CM represents quantities measured with respect to the clump's center of mass. In this way, the SAM will be simply given by $\mathbf{J}/M_{\text{gas}}$.

It is worth noting that, in some cases, while tracking a clump over time, its member SPH particles can cross the periodic boundaries of the numerical box, appearing at the other side. Failing to take this into account greatly increases the measured SAM and radius of the clump. To avoid this problem, for each clump we perform a coordinate translation to move it to the center of the box. This avoids the problem because no clump is large enough nor moves fast enough to touch the boundary at any time during its evolution when placed at the simulation center at the time it is defined.

\section[]{Results}
\label{sec:results}

In this section we now investigate several aspects of the distribution and evolution of the AM. We first investigate the instantaneous distribution of the numerical clump sample in the \jR\ diagram, combining data from four different snapshots. Next, in order to understand the redistribution of the AM during fragmentation, we follow the evolution of the SAM of particle sets either as  ``lagrangian sets'' (i.e., consisting of the same set of SPH particles at all times) or as connected regions above a threshold at all times. The tracking over time is done either from the past or towards the future of the time $\tdef$ at which the clumps are defined as connected regions above a threshold.


\subsection{\jR\ relation for the numerical clump sample at fixed times}
\label{subsec:fixed times}

\begin{figure}
 \includegraphics[width=\columnwidth]{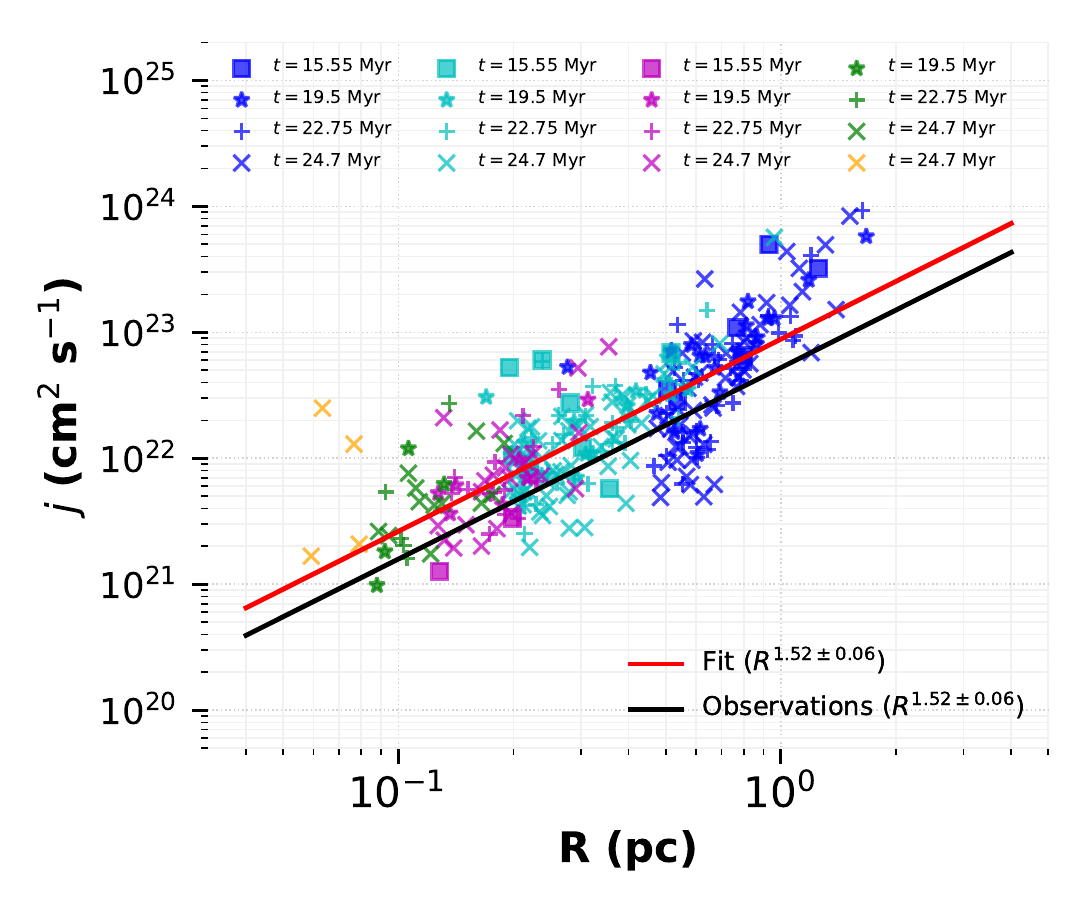} 
 \caption{The \jR\ relation for the numerical clump sample, considering clumps defined as connected regions above a threshold density $\nth$, at times $t=17.92$, $t=19.92$, $t=23.24$ and $t=25.23$ Myr. The black line represents the fit to the observations compiled in Fig.\ \ref{fig:data}, while the red line is the fit for the simulation clump sample. The symbols are colored according to the density threshold used to define the clumps: $n_{\text{th}} = 10^{3}$ (blue), $3 \times 10^{3}$ (cyan), $10^{4}$ (magenta), $3 \times 10^{4}$ (green), and $10^{5}$ cm${}^{-3}$ (yellow). Clumps meet the condition $\Sigma > 30$ M${}_{\odot}$pc${}^{-2}$. The different symbols represent the time at which the clumps were defined, as given in the figure legend. The clumps in the numerical sample clumps are seen to exhibit a trend similar to that of the observations. }
 \label{fig:fixed times}
\end{figure} 

Figure \ref{fig:fixed times} shows the \jR\ relation for the numerical clump sample, with the radius $R$ calculated from the clump's volume according to eq.\ (\ref{eq:total vol}). In this figure, the color code corresponds to the density threshold used to define clumps, and the different symbols denote the clump definition time, $\tdef$. The red line represents the fit to the data for the numerical sample, given by 
\begin{equation}
    j = 10^{22.9 \pm 0.03} \left(\frac{R}{1 {\rm pc}}\right)^{1.52 \pm 0.06} {\rm cm}^2~ {\rm s}^{-1},
    \label{eq:numerical fit}
\end{equation}
while the black line shows the fit to the observational sample shown in Fig. \ref{fig:data}. It can be seen that the numerical sample 
exhibits a slope and intercept remarkably close to those of the observational sample. This suggests that the GHC simulation adequately represents the AM redistribution processes taking place in actual clouds and clumps.



\subsection{Specific angular momentum evolution of lagrangian particle sets}
\label{subsec:lagrangian}

We now discuss the temporal evolution of the SAM of a few lagrangian particle sets, each of which constituted a connected clump at the time $\tdef$ when they were defined. The particles are tagged, and we follow them as they advance either from the past ($t < \tdef$) or to the future ($t > \tdef$). When followed from the past, we then present the forward evolution starting from the earliest time reached, and finishing at the definition time $\tdef$. When followed to the future, we also show the forward evolution {\it starting} from $\tdef$.

At every time during their evolution, we compute the AM of the sets of particles according to eq.\ (\ref{eq:ang_mom}), and their size according to eq.\ (\ref{eq:geometric average}), recalling that, at times $t \ne \tdef$, the particles in general do not constitute a connected set, and therefore the size of the region they occupy cannot be computed from the sum of their individual volumes as in eq.\ (\ref{eq:total vol}), and instead, eq.\ (\ref{eq:geometric average}) must be used (cf.\ Sec.\ \ref{subsec:radius}).



\subsubsection{Tracking nested lagrangian particle sets from the past}
\label{subsubsec: same cloud}

\begin{figure*}
 \includegraphics[width=18cm, height=8cm]{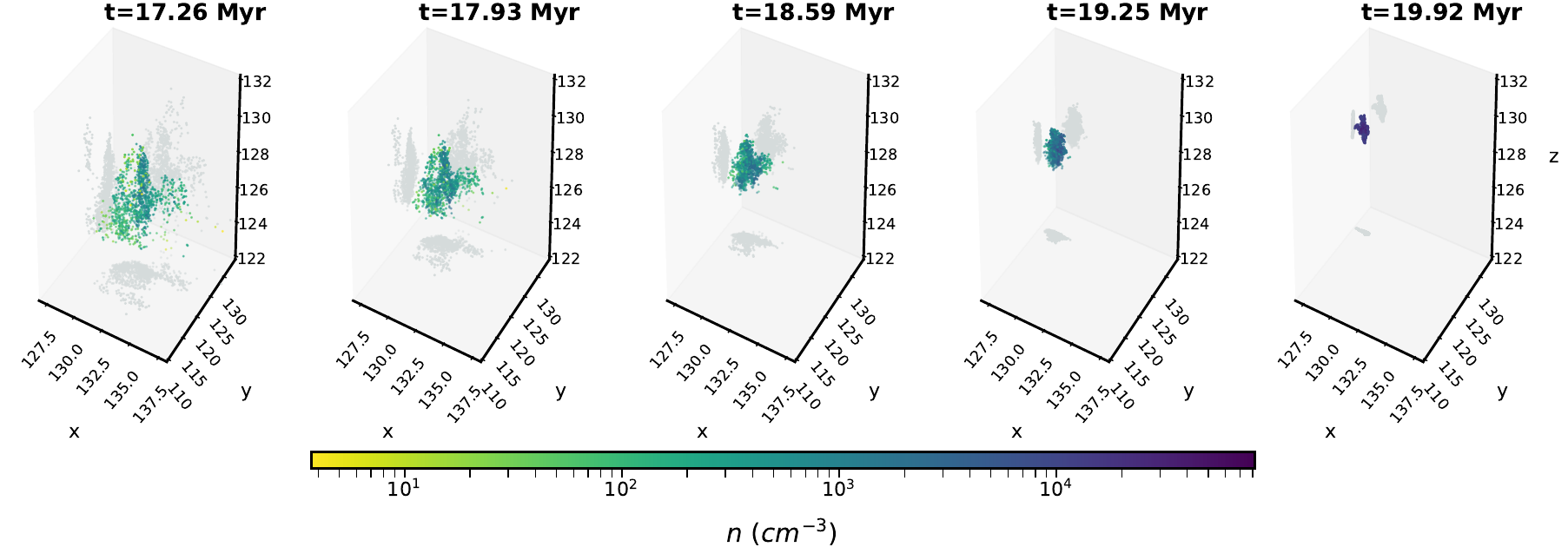}
 \caption{Past evolution of the density and spatial distribution of the lagrangian set of particles making up clump C3 (see Table \ref{tab:all clumps}), defined at $\tdef =  19.92$ Myr (rightmost image) and $n_{\rm th} = 10^{4}\, \pcc$. The total time span is 2.66 Myr. The set of particles is seen to become denser by nearly two orders of magnitude and to contract from a size of several parsecs to $\sim 1$ pc. The colored dots show the SPH particles, with their color indicating their density. The grey dots indicate their projection on the bounding planes.}
 \label{fig:muestra one}
\end{figure*}

\begin{figure}
  \centering
\includegraphics[width=0.5\textwidth]{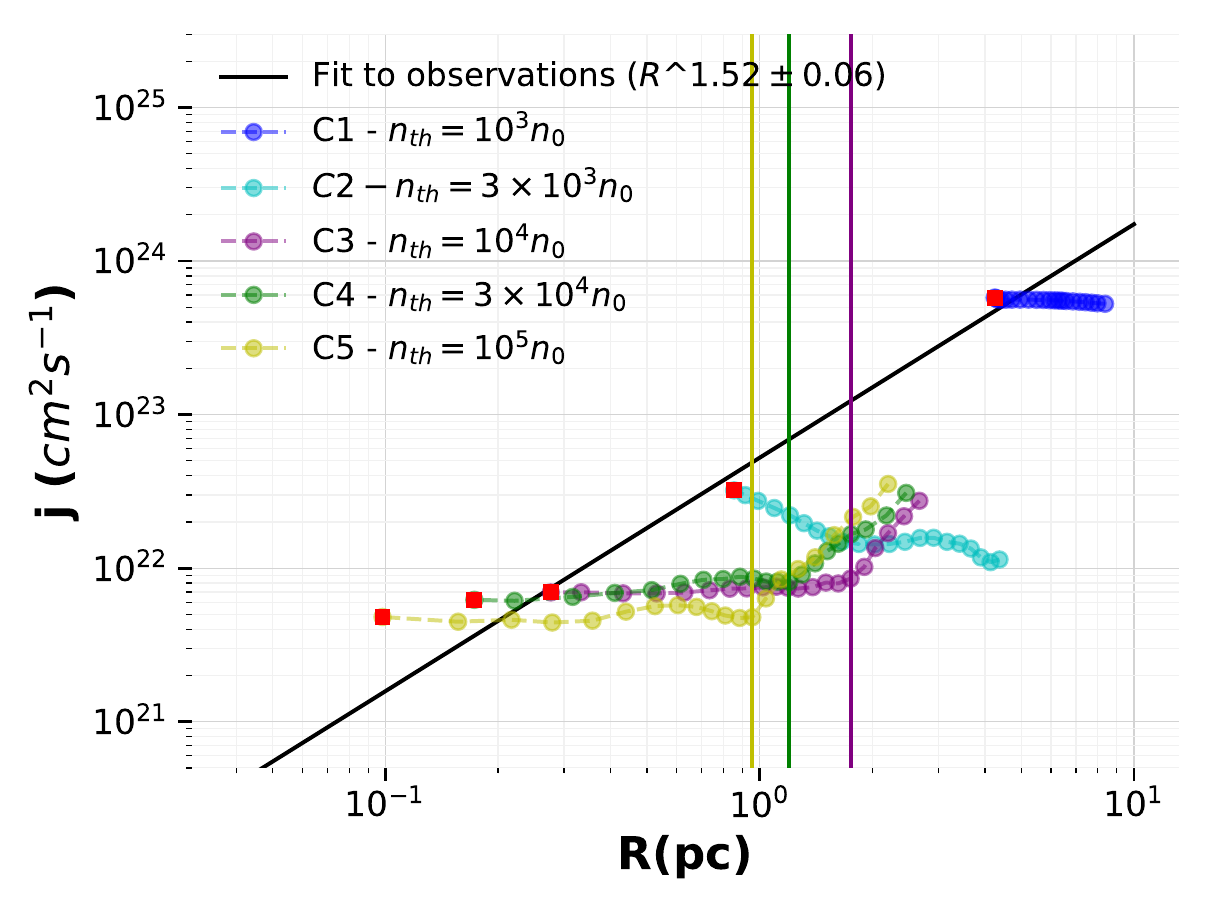}
  \caption{Evolutionary tracks from the past (since $t=17.27$ Myr) up to $\tdef = 19.92$ Myr for the five lagrangian sets C1-C5, which make up the nested clumps shown in Fig.\ \ref{fig:muestra} at $\tdef$. The lagrangian sets evolve from right to left in this diagram, with the red filled square representing the final time, $\tdef$. The vertical lines mark the radius of each set at which its track changes slope, with their color indicating their corresponding lagrangian set. The black line represents the fit made to the observational sample shown in Fig.\ \ref{fig:data}. For the three smallest clumps, two main periods of evolution can be identified: an early period with an evolution along a slope similar to the observed \jR\ relation, and a late period with $j$ approximately constant. For the two largest clumps, only the $j \sim$ cst.\ stage is observed.}
  \label{fig:evo same clump}
\end{figure}

We consider the five hierarchically-nested clumps shown in Fig.\ \ref{fig:muestra} (clumps C1-C5 in Table \ref{tab:all clumps}), which were defined with five different density thresholds at time $\tdef = 19.92$ Myr, and we track them from the past---i.e., from $t < \tdef$---as lagrangian sets . The tracking was performed over a period of $2.65$ Myr. Figure \ref{fig:muestra one} thus shows the evolution of the clump defined at threshold $n_{th} = 10^{4} \pcc$ from $t = \tdef - 2.65 = 17.27$ Myr to $\tdef$. It can be seen that, at the earliest time, the set of particles was nearly two orders of magnitude less dense and several times more extended. 

The evolution of each lagrangian set in the \jR\ diagram over the 2.65 Myr prior to $\tdef$ is shown in Fig. \ref{fig:evo same clump}. In this figure, the clumps evolve from right to left, as they shrink and become denser; the colors correspond to different density thresholds, and the vertical lines indicate the time at which a change in the slope of the evolutionary track occurs for the lagrangian set of the corresponding color. The red filled square at the extreme left of each track corresponds to $\tdef$, which in this case is the {\it final} time. The black line represents the fit to the observational data from Fig. \ref{fig:data}. 


For the three smallest particle sets, two main periods of evolution can be identified: an early period in which the evolutionary track has a slope similar to that of the observational relation, and a late one over which $j$ remains approximately constant. The transition to $j \sim $ cst. occurs when the clump is already very compact. The period of AM loss occurs when the lagrangian set of particles is very scattered, and presumably with many ``intruders''. For the two largest clumps, defined at the lowest density thresholds, $j \sim$ cst. over the entire tracking period.

The above result leads us to suggest that the change in slope observed in Fig.\ \ref{fig:evo same clump} may be due to the more scattered state of the clump member particles in the past, with a larger number of intruder particles interspersed among them. The member particles can then transfer their AM to the intruders, and be able to contract. Instead, at later times, the member particles form a denser, more connected ensemble without many partners to exchange their AM with, and thus tend to conserve it.

\begin{figure}
 \includegraphics[width=\columnwidth]{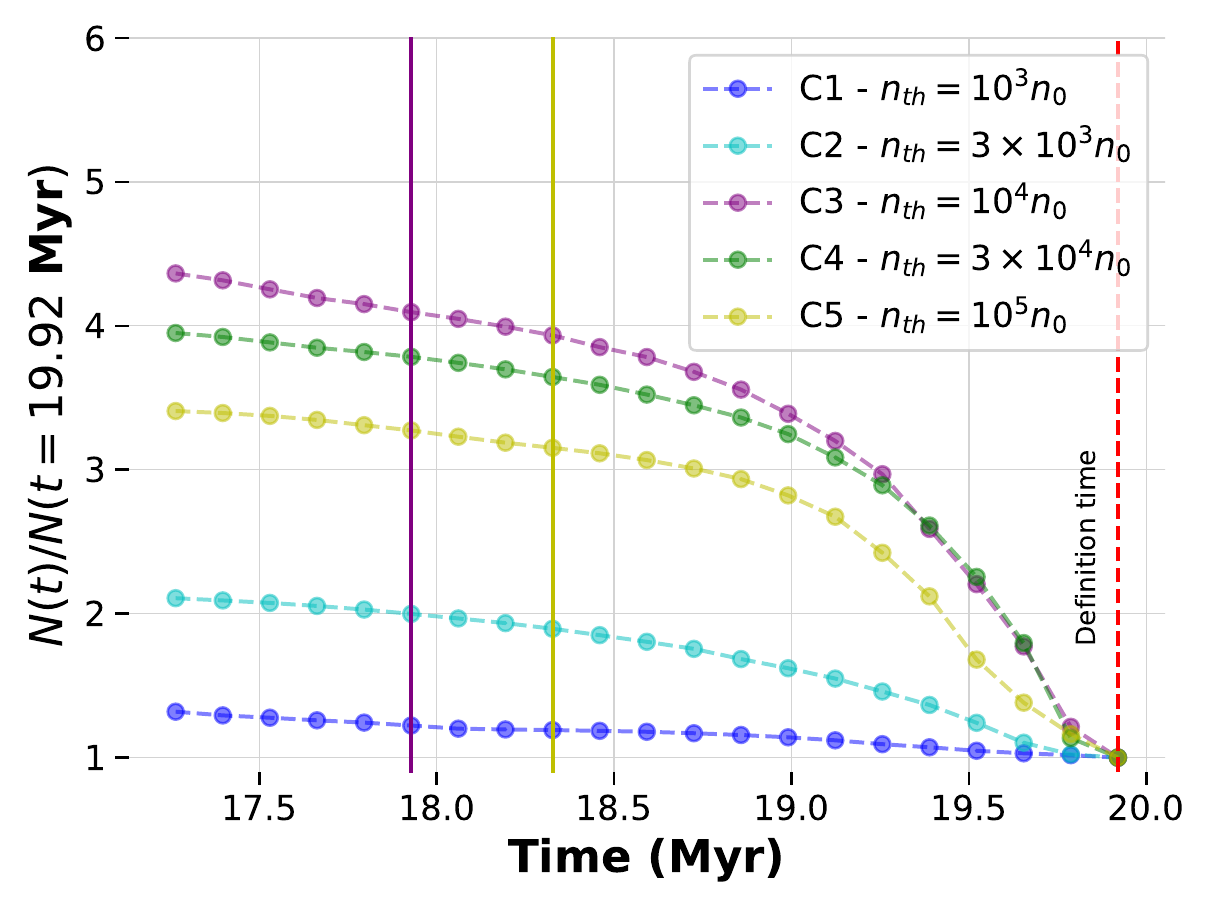}
 \caption{Ratio of the number of intruder particles within the minimal rectangular box that encloses each of the lagrangian sets C1-C5 at the indicated time $t$ to the number of intruders at $t = \tdef$ (denoted by the red vertical dotted line). The solid vertical lines represent the time at which the slope of the corresponding evolutionary track in Fig.\ \ref{fig:evo same clump} changes, and their color indicates the corresponding clump. Note that the green vertical line is coincident with the vertical yellow line, and so it is not visible. }
 \label{fig:extra particles mass}
\end{figure}

In order to test this hypothesis, we define a ``minimal rectangular box'' as a rectangular volume enclosing the lagrangian set at each time, whose sides are placed at the coordinates of the most extreme particles along each coordinate axis (as in Fig. \ref{fig:box}), and determine the fraction of intruder particles in this volume. 

It is important to note that, if a lagrangian set has significant protrusions in several directions, it will define a very large volume, which will have a high percentage of intruder particles. To minimize the impact of this geometric effect, we consider the ratio of the number of intruder particles within the minimal box at time $t$, $N(t)$, to the number of intruders within the corresponding minimal box at the final time ($N(\tdef)$. Figure \ref{fig:extra particles mass} shows the evolution of this ratio for the five lagrangian sets considered. The colored vertical lines represent the time at which the slope of the evolutionary track for each clump in Fig.\ \ref{fig:evo same clump} changes, if it does. The green line coincides with the yellow line, and thus it is not visible. The vertical red dotted line indicated the definition time, $\tdef$.

Figure \ref{fig:extra particles mass} shows that, on average, the particle sets exhibiting an evolutionary track segment parallel to the observational slope in the \jR\ diagram (yellow, green and purple lines) had in the past few Myr a normalized intruder fraction larger than three times the value at the final time. Instead, the clumps that do not exhibit such stage (blue and cyan lines) had an initial value of this ratio at most twice the final value. This result supports the hypothesis that a lagrangian set of SPH particles can lose AM as long as it has a companion set of particles to transfer it to, and instead evolves at roughly constant $j$ when it is evolving mostly as an isolated entity. Also, this argues against the dominant torques being gravitational, since these should act over long distances, and thus should not require the companions to be nearby. They are also not expected to be important when the densities of the interacting fluid parcels are similar.



\subsubsection{Tracking independent lagrangian particle sets from the past}

\begin{figure}
 \includegraphics[width=\columnwidth]{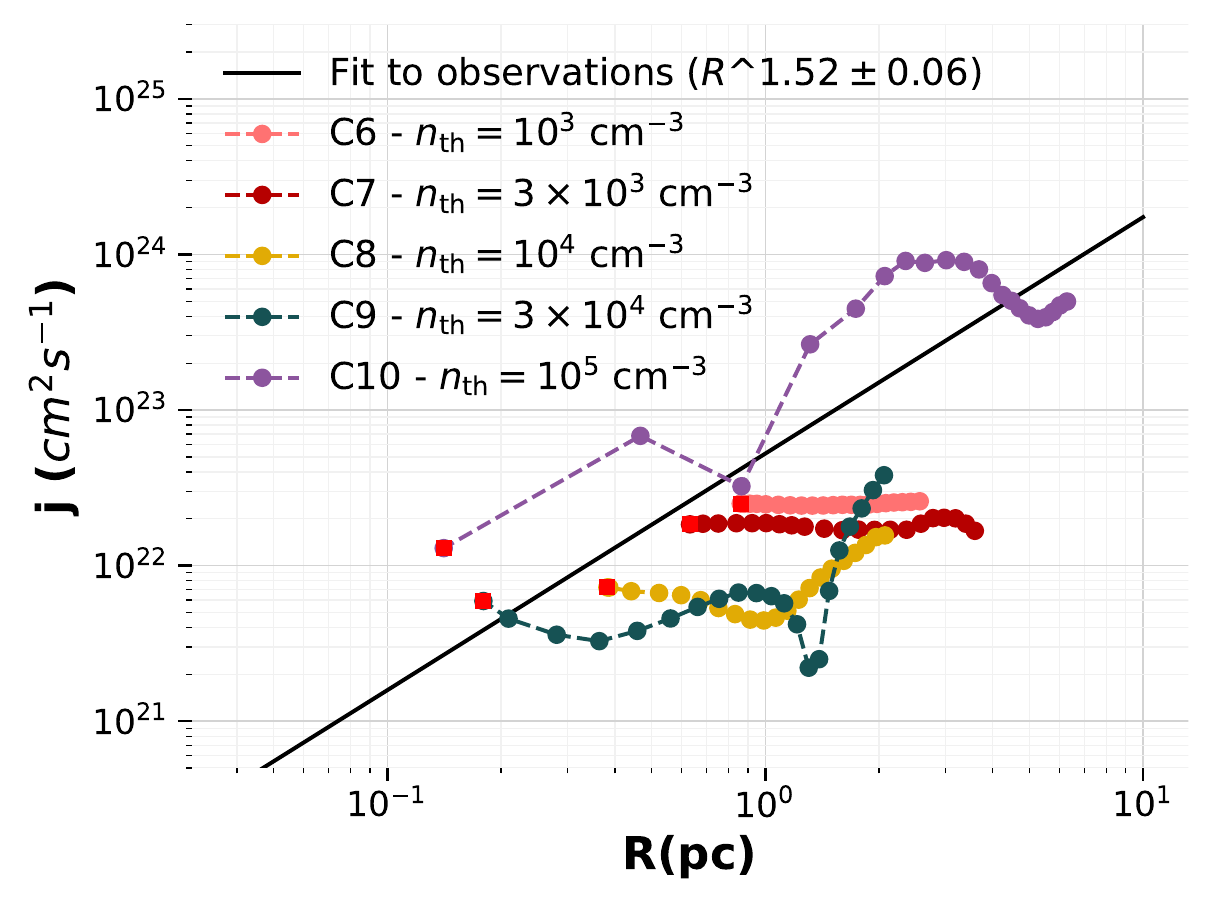}
 \caption{Evolution of the lagrangian sets C6-C10, defined in different regions of the numerical box with a density threshold $n_{\rm th} = 3 \times 10^{3}\, \pcc$ at time $\tdef = 25.23$ Myr (red filled squares), and followed to earlier times by $2.7$ Myr. The direction of evolution is therefore from right to left. The black line represents the fit made to the observational sample shown in Fig. \ref{fig:data}.}
 \label{fig:evo same dens}
\end{figure}

In the previous subsection we followed the evolution in the \jR\ diagram of the lagrangian sets corresponding to five hierarchically-nested clumps originally defined with various threshold densities within the same global density enhancement. The fact that they hierarchically nested raises the question of whether the observed two-stage evolution might be a peculiarity of the chosen parent clump, and so here we track the (past) evolution of five lagrangian sets (clumps C6-C10 in Table \ref{tab:all clumps}), originally defined as independent clumps at various random locations in the simulation using the same five density thresholds as in Fig.\ \ref{fig:evo same clump}, and at definition time of $\tdef = 25.23$ Myr.

The evolution of these five lagrangian particle sets during the 2.7 Myr prior to $\tdef$ is shown in Fig.\ \ref{fig:evo same dens}. It can be seen that the behavior of these clumps is similar to those of Fig.\ \ref{fig:evo same clump}, since the change in slope in the \jR\ diagram occurs for thresholds $\gtrsim 10^{4}\, \pcc$. This supports the conclusion that the behavior observed in Fig.\ \ref{fig:evo same clump} is not a special feature of the parent structure of the particle sets shown there.


\subsubsection{Tracking lagrangian sets at different density thresholds to the future}
\label{subsubsec:future}

\begin{figure*}
\centering
 \includegraphics[width=12.5cm]{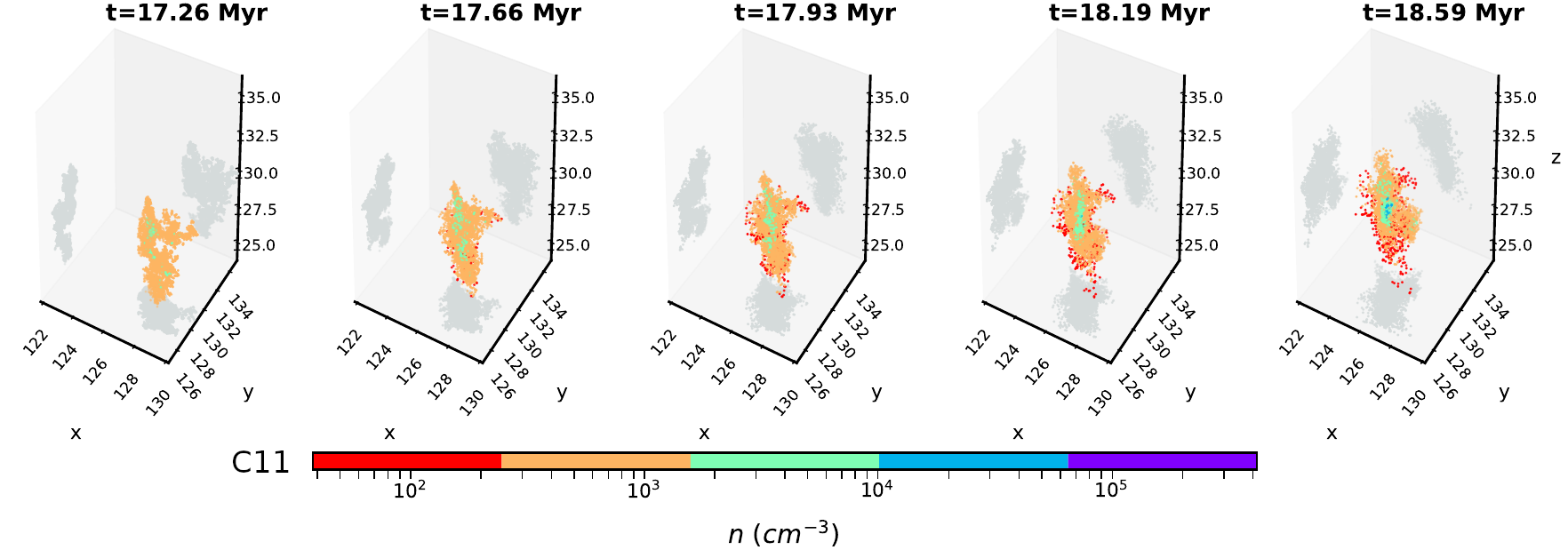}\\
 \includegraphics[width=12.5cm]{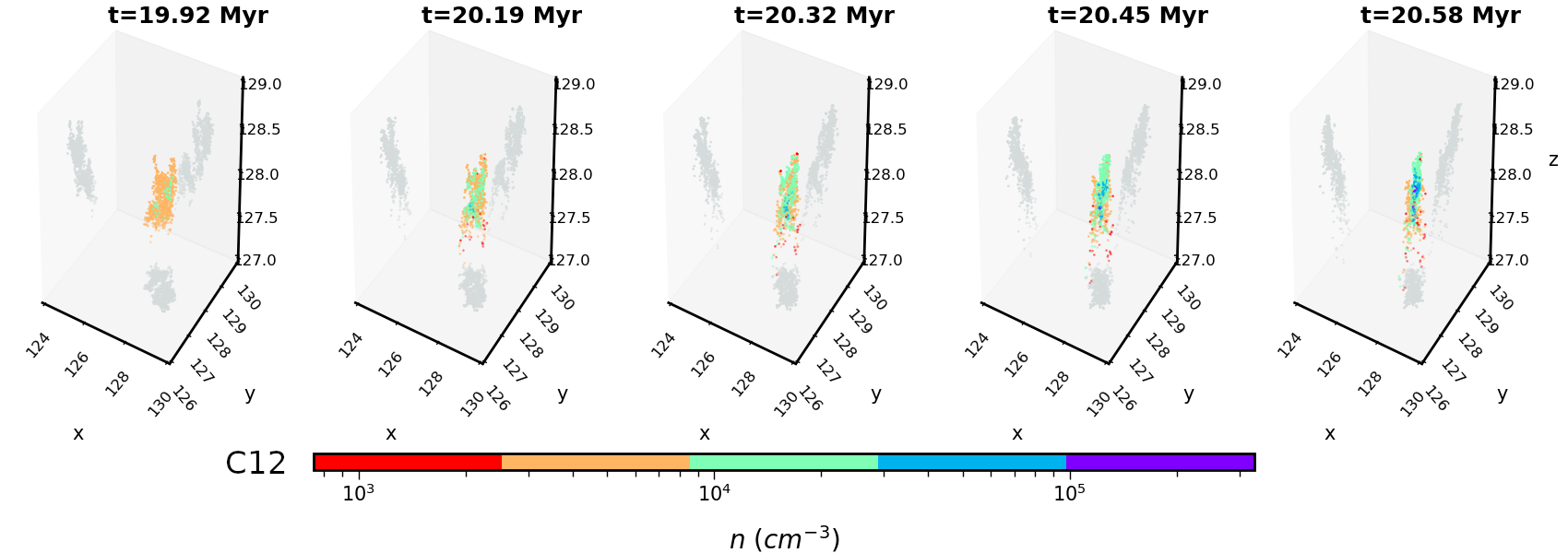}\\
 \includegraphics[width=12.5cm]{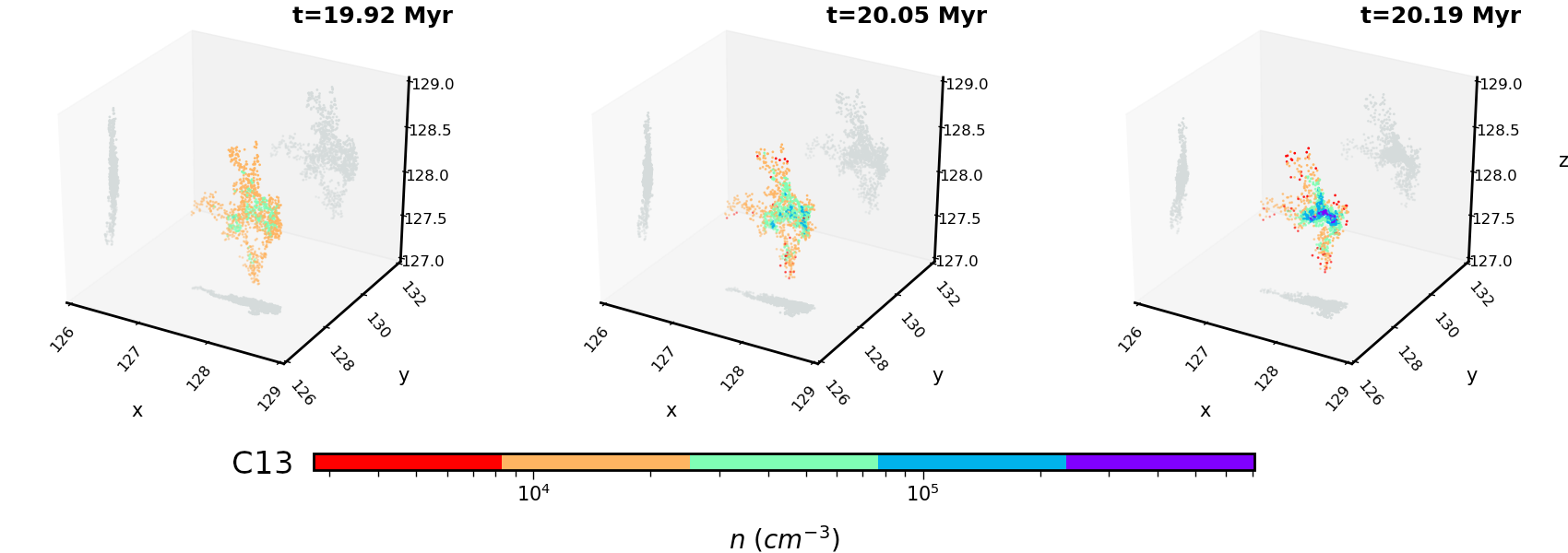}\\
 \includegraphics[width=12.5cm]{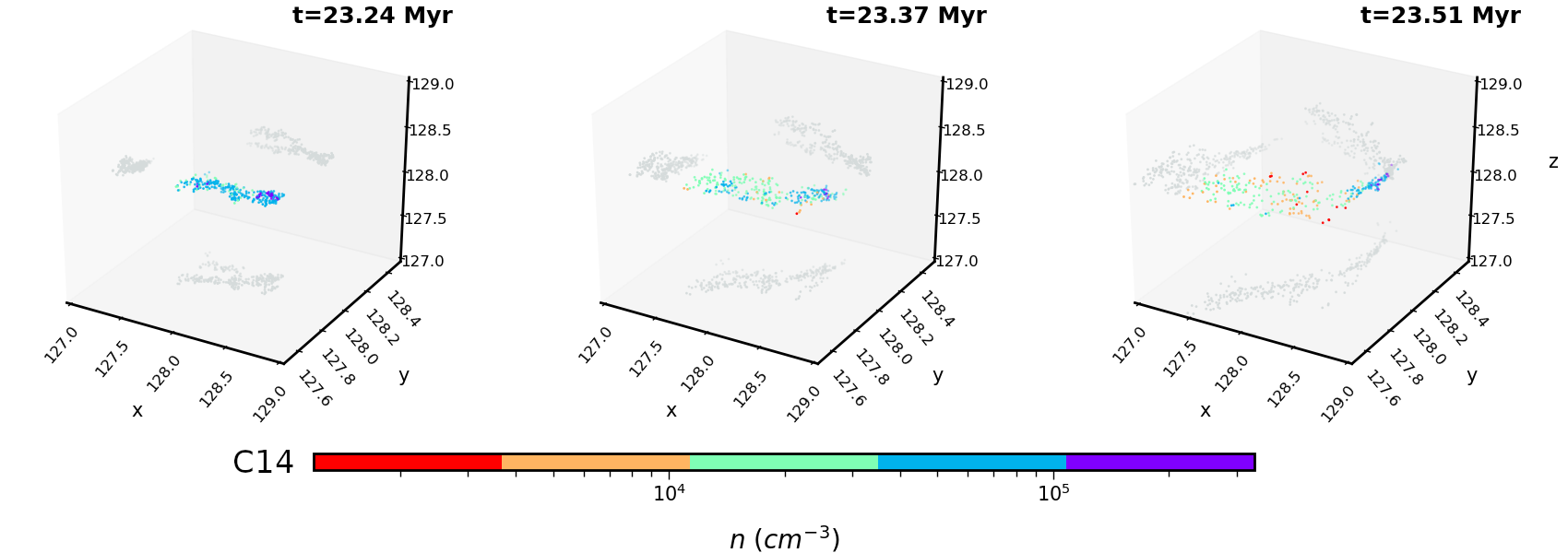}\\
 \includegraphics[width=12.5cm]{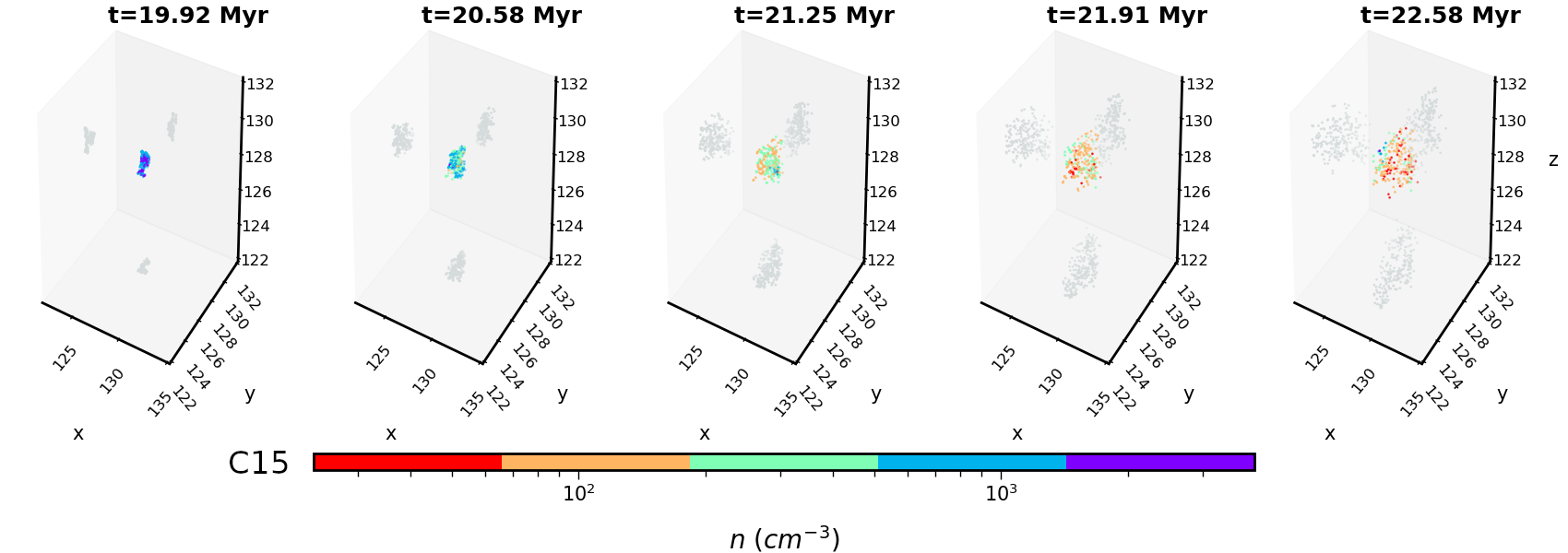}
 \caption{Density and spatial distribution evolution of the set of member particles of five clumps defined by density thresholds $n_{\rm th} = 10^{3}\, \pcc$ (C11), $n_{\rm th} = 3 \times 10^{3}\, \pcc$ (C12), $n_{\rm th} = 10^{4}\, \pcc$ (C13), $n_{\rm th} = 3 \times 10^{4}\, \pcc$ (C14), and $n_{\rm th}= 10^{3}\, \pcc$ (C15), at times $\tdef = 17.26$ Myr (C11), $\tdef = 19.92$ Myr (C12 and C13), and $\tdef = 23.24$ Myr (C14 and C15). These clumps were tracked toward the future until they formed sinks. It can be seen that, for Clumps A-D, the innermost part of the particle set undergoes collapse, while the outermost particles decrease their density and appear to escape from the clump. Instead, Clump E turns out to be a transient clump that disperses as time passes.}
 \label{fig:muestra one future}
\end{figure*}

We now consider the evolution {\it towards the future} of five lagragian sets of particles defined by density thresholds $n_{\rm th} = 10^{3}\, \pcc$ (C11 in Fig.\ \ref{fig:muestra one future}), $n_{\rm th} = 3 \times 10^{3}\, \pcc$ (C12), $n_{\rm th} = 10^{4}\, \pcc$ (C13), $n_{\rm th} = 3 \times 10^{4}\, \pcc$ (C14) and $n_{\rm th}= 10^{3}\, \pcc$ (C15). C11 was defined at time $\tdef = 17.2$ Myr, C12, C13 at $\tdef = 19.92$ Myr, and C15, C14 at $\tdef = 23.24$ Myr.


The spatial distribution and density evolution for all clumps is shown in Fig. \ref{fig:muestra one future}. It can be seen that, as time proceeds, some of the set member particles of clumps C11 to C14 appear to {\it disperse} (red dots), while another group of particles proceeds to collapse, becoming denser and much more compact (cyan, blue and purple dots). Regarding clump C15, it is seen that the entire set of particles disperses, indicating that this is an example of a transient, dispersing clump.

\begin{figure}
\centering
\includegraphics[width=\columnwidth]{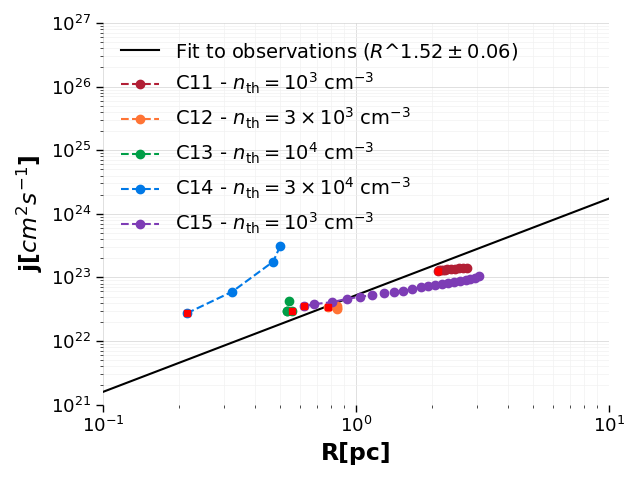}
 \caption{Evolution of the specific angular momentum of the five lagrangian sets of particles shown in Fig. \ref{fig:muestra one future}. These clumps were tracked toward the future until they formed sinks.  The direction of evolution in the figure is from left to right, with the red filled squares denoting $\tdef$. The black line represents the fit to the observational data shown in Fig. \ref{fig:data}.}
 \label{fig:evo to future}
\end{figure}

In Fig.\ \ref{fig:evo to future} we show the evolutionary tracks in the \jR\ diagram for these five clumps. In these tracks, the direction of time evolution is from left to right. Note, however, that, because of the tracking to the future, in this case the formation of sink particles among the set members cannot be prevented---in the cases of tracking to the past, the absence of sinks at $\tdef$ guaranteed that no sinks were present also during the previous evolution towards $\tdef$---, and indeed a sink appears after 1.32 Myr in C11. Once a sink appears and accretes several of the SPH particles, it absorbs part of the AM of the set. Since we are interested only in the AM exchanges between fluid particles, the evolutionary tracks in this figure are limited to the time before a sink appears.

Figure \ref{fig:evo to future} shows that none of the evolutionary tracks undergo an abrupt slope change like those seen in Fig.\ \ref{fig:evo same clump}, nor exhibit a period of evolution at constant $j$. This suggests that the dynamics of these sets of particles is always dominated by interactions with intruder particles. 


More important, however, is the observation that a fraction of the member particles reduces its density to values below the definition threshold and recedes from the clump, in spite of having been initially all above the threshold density. We have observed this phenomenon in all sets of particles tracked to the future that develop a collapse center. {\it This suggests that losing a fraction of the clump's mass to carry away the AM is a necessary condition for the rest of the clump to be able to collapse,} in a manner similar to the process of disk accretion towards a central object. In the latter case, it is well known that some orbital AM must be transferred outwards in order for the rest of the  material to be able to flow towards the center. It is important to note that this mechanism of AM transfer does not require the presence of a magnetic field to operate, and instead operates directly among neighboring fluid particles, probably via ram pressure (or eddy) torques (first term on the right-hand side of eq.\ [\ref{eq:total torque}]) among them.

\subsection{Specific angular momentum evolution of a regular clump (always defined as a connected overdensity)}
\label{subsec:variable clump}

\begin{figure*}
 \includegraphics[width=18cm, height=8cm]{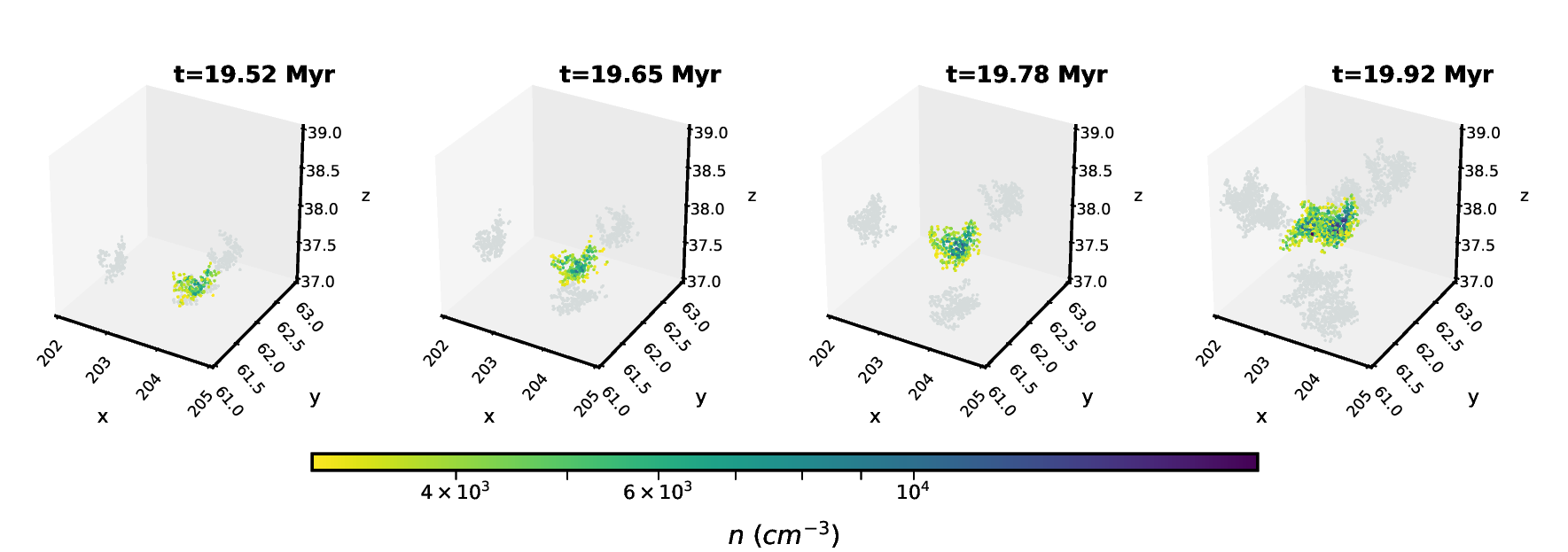}
 \caption{Spatial evolution of clump C16, defined as a connected object above $n_{\rm th} = 3 \times 10^{3}\, \pcc $, over a period of $0.4$ Myr starting at $\tdef = 19.52$ Myr. In this case, the clump-finding algorithm was used to define the clump at all times over the region denoted by the boxes. Due to this form of definition, the clump does not consist of the same fluid particles at all times, but rather grows in mass and size due to accretion from its environment, while shedding some of its other particles, in spite of harboring a local process of gravitational collapse.}
 \label{fig:muestra same tresh}
\end{figure*}

\begin{figure}
 \includegraphics[width=\columnwidth]{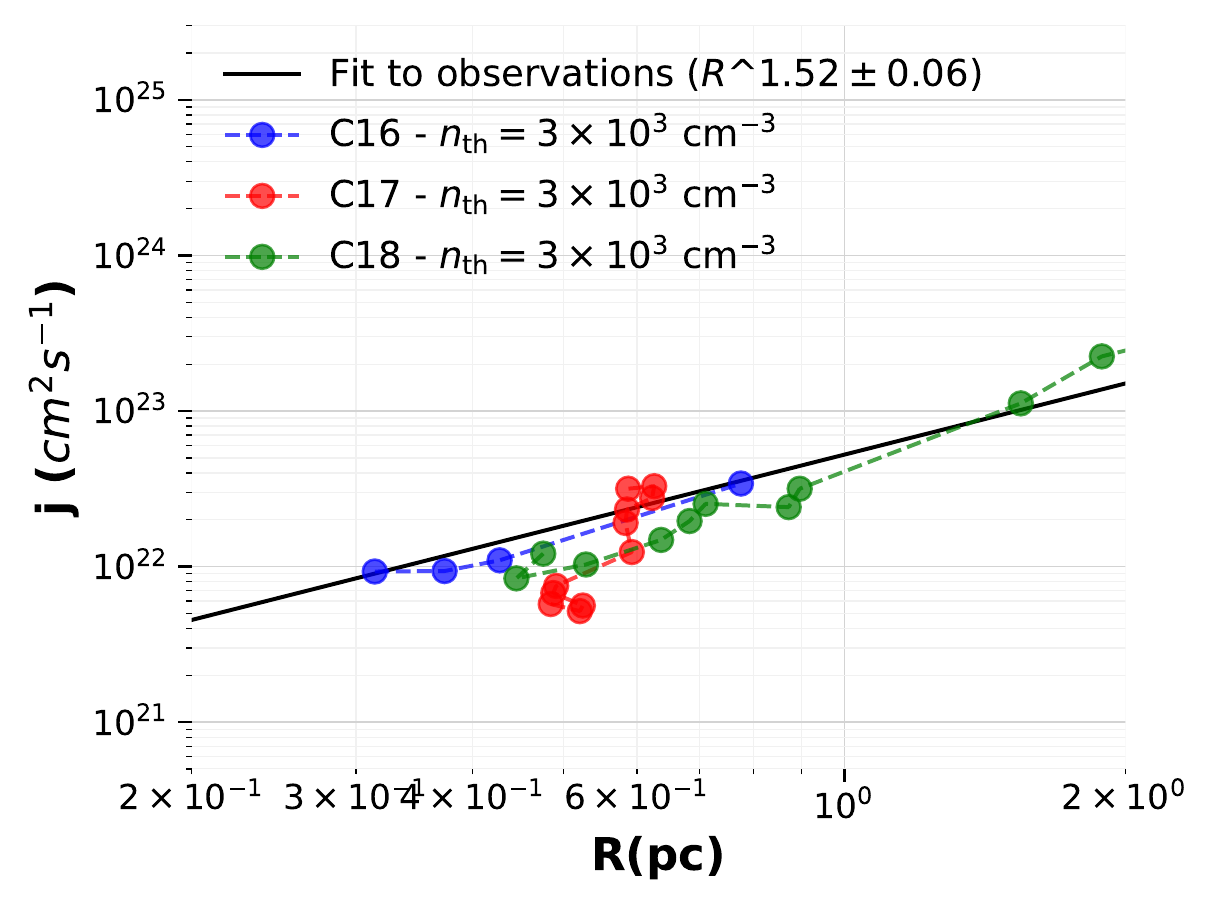}
\caption{Specific angular momentum evolution of clumps C16-C18 (C16 is shown in Fig.\ \ref{fig:muestra same tresh}), defined at times $t=17.26$ (C16) and $17.93$ (C17 and C18) as connected regions above $n_{\rm th} = 3 \times 10^{3}\, \pcc$, each one tracked until it forms sinks. The black line represents the slope of the fit to the observational data shown in Fig.\ \ref{fig:data}. As time advances, clumps grow in size and become more massive, and their SAM increases as well, although they evolve near the observational slope over the whole period}.
 \label{fig:evo same tresh}
\end{figure}

We now consider the evolution of clumps defined in the traditional way (clumps C16-C18 in Table \ref{tab:all clumps}); that is, as connected sets of particles above a density threshold throughout their evolution. That is, in this case, the set of SPH particles that compose it is not lagrangian. Instead, the clumps continually exchange fluid particles with their environment. Although both accretion and particle loss occur, the former dominates, and so the clumps grow in mass, size and mean density. Specifically, the clumps are defined at a threshold $\nth = 3 \times 10^3\, \pcc$ at times $t=17.26$ (C16) and $17.93$ Myr (for C17 and C18). As an illustration, Fig.\ \ref{fig:muestra same tresh} shows the evolution of the spatial distribution, density, and mass of C16. 

In Fig.\ \ref{fig:evo same tresh} we show the evolution of these three clumps in the \jR\ diagram while they grow in size and mass. For all three of them, the increase in both radius and SAM appears to occur along evolutionary tracks close to the locus of the observational sample in this diagram. 


It is important to note that this way of following the clumps is similar to how they would be observed in practice, for example through the emission of a tracer with some specific effective or critical excitation density \citep[e.g.,] [] {Shirley15}, which allows observation only above a threshold. This suggests that the clumps that make up observational samples, such as the one shown in Fig.\ \ref{fig:data}, do not correspond to a sequence of objects that have contracted coherently as a single unit, but rather undergo a true fragmentation process such as that observed in Fig.\ \ref{fig:evo to future}, in which one part of the clump is lost, carrying AM with it, and allowing the remainder to contract.



\section{Discussion and implications}
\label{sec:discussion}

\begin{figure*} 
  \centering
  \includegraphics[width=0.3\textwidth]{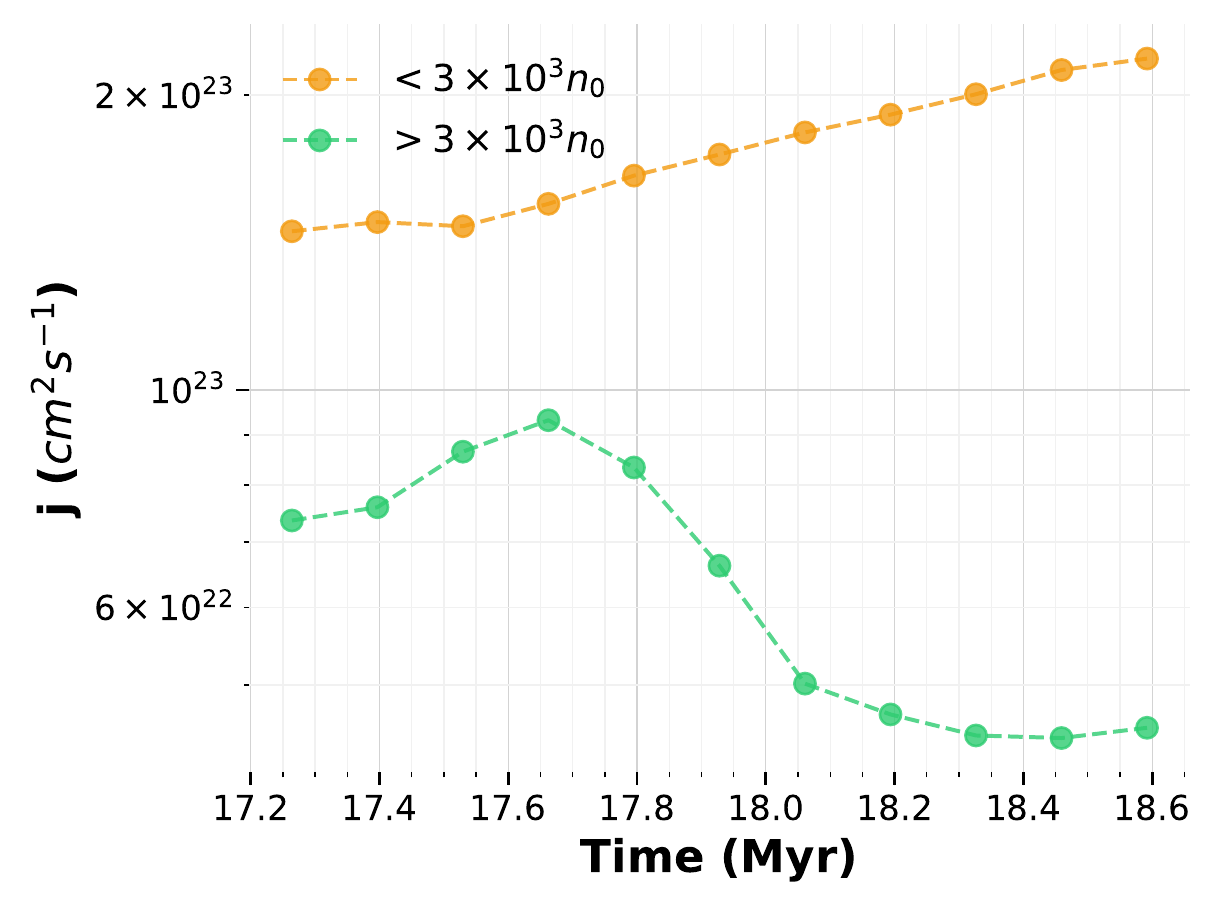}
  \includegraphics[width=0.3\textwidth]{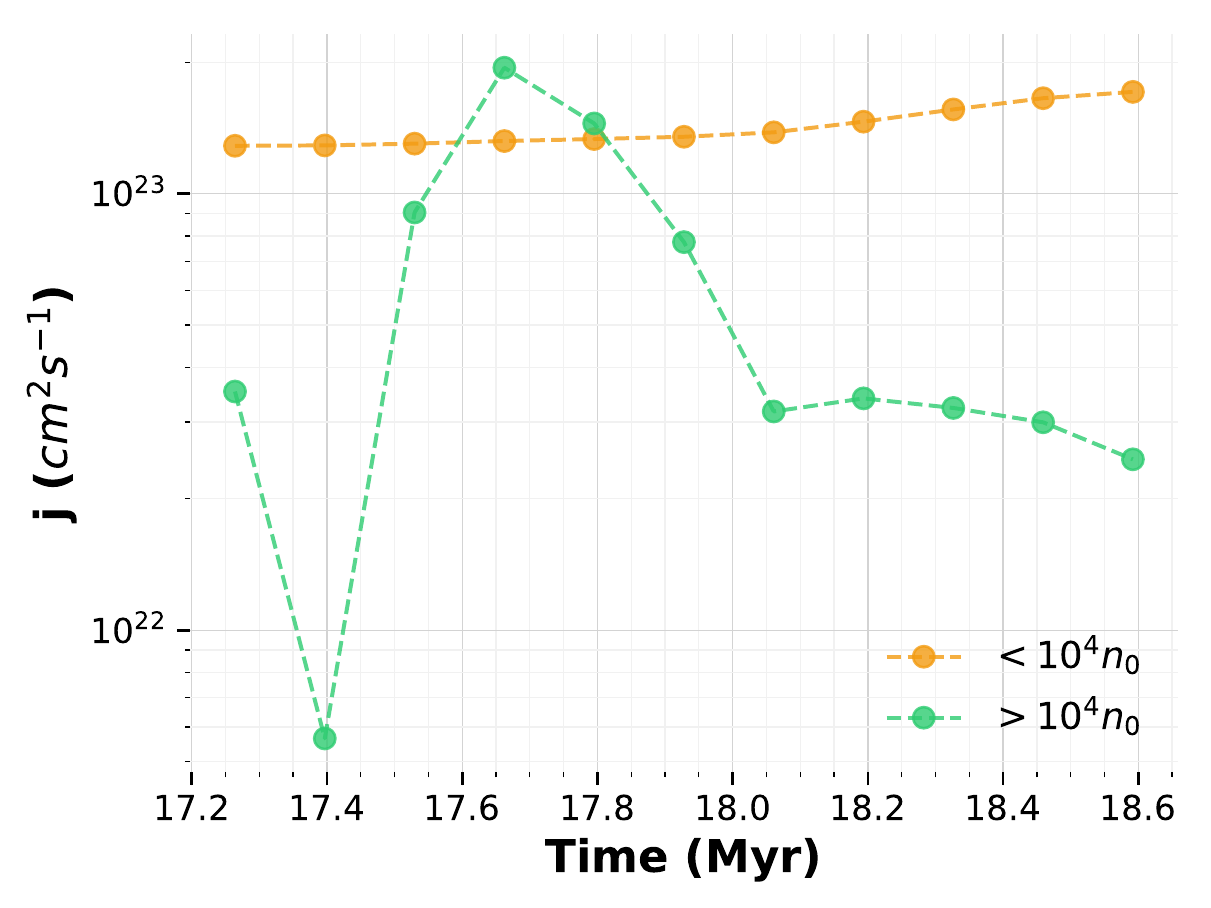}
  \includegraphics[width=0.3\textwidth]{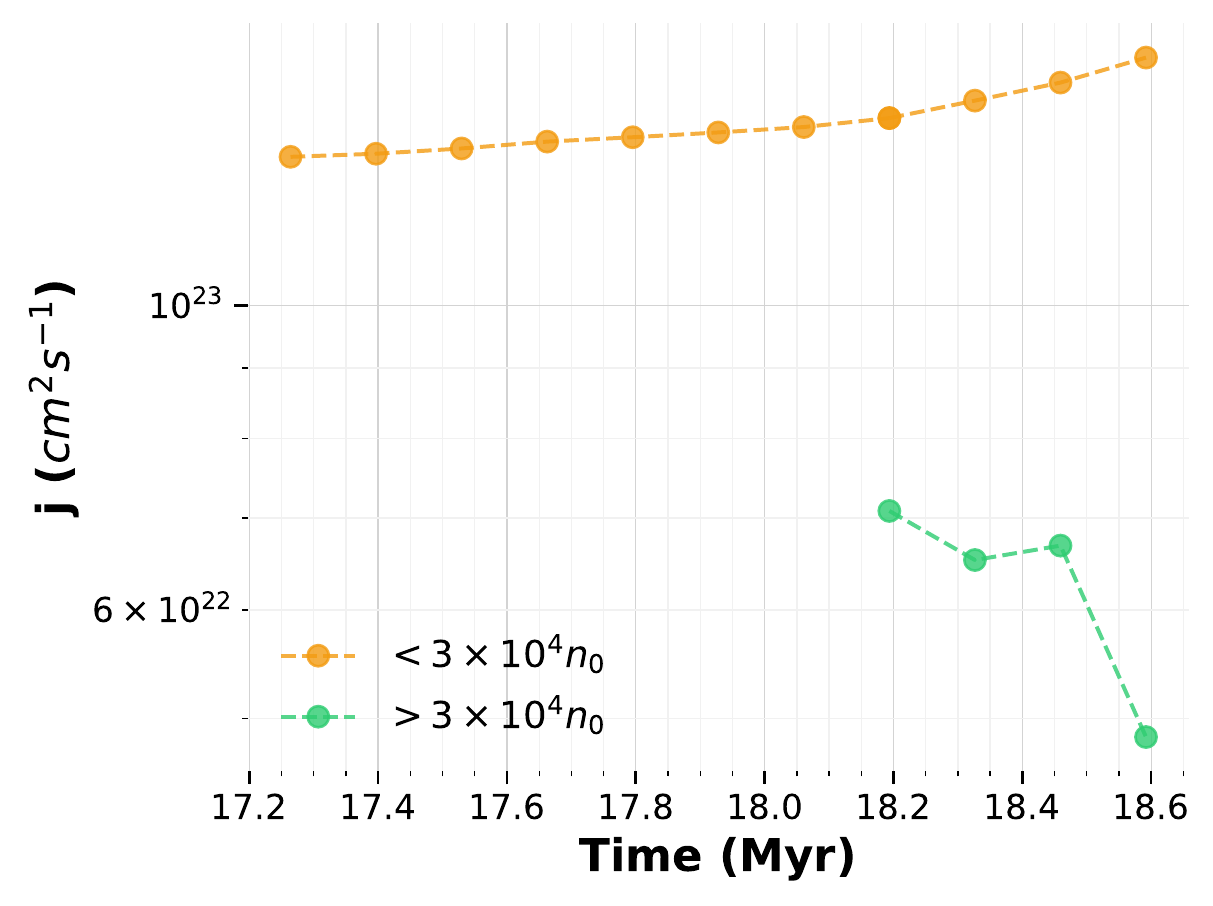}
  \hfill
  \includegraphics[width=0.3\textwidth]{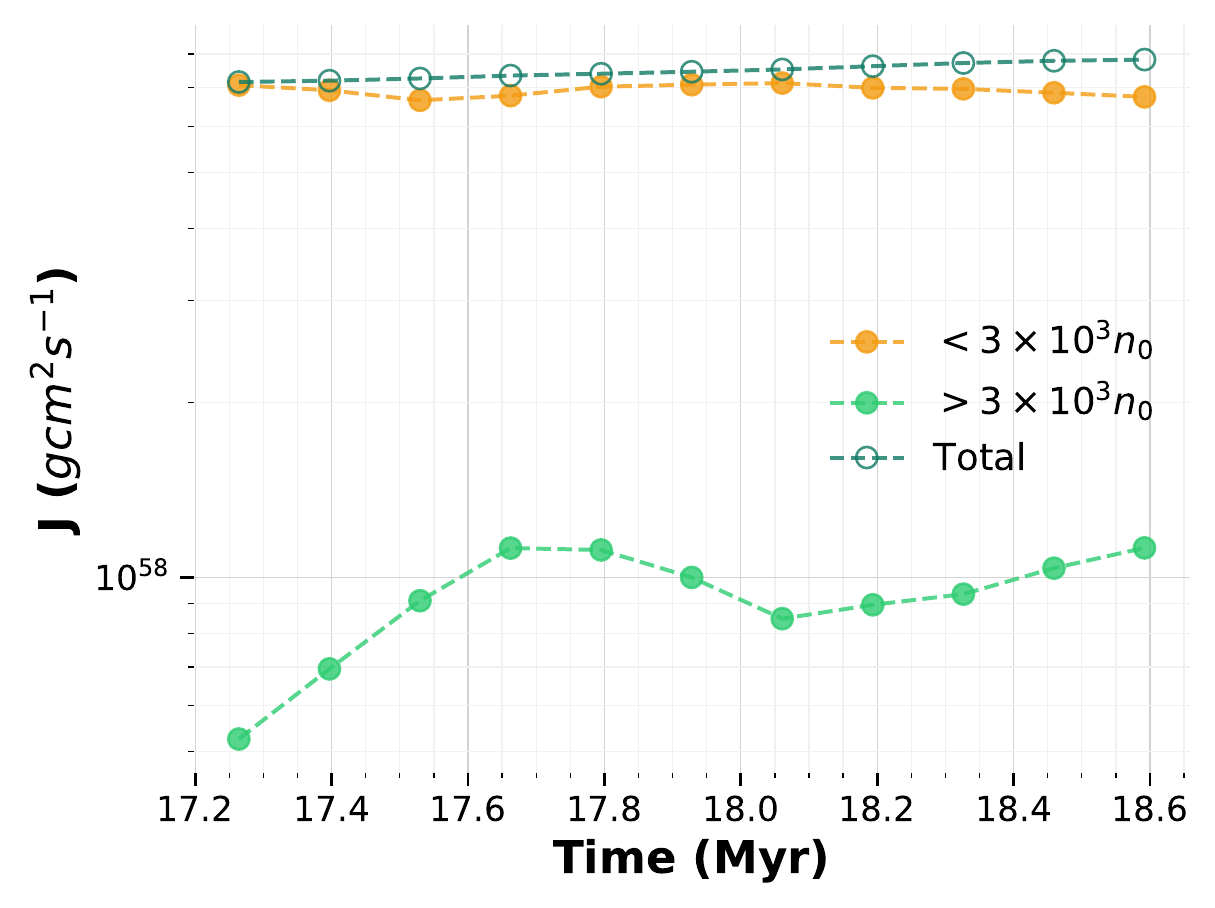}
  \includegraphics[width=0.3\textwidth]{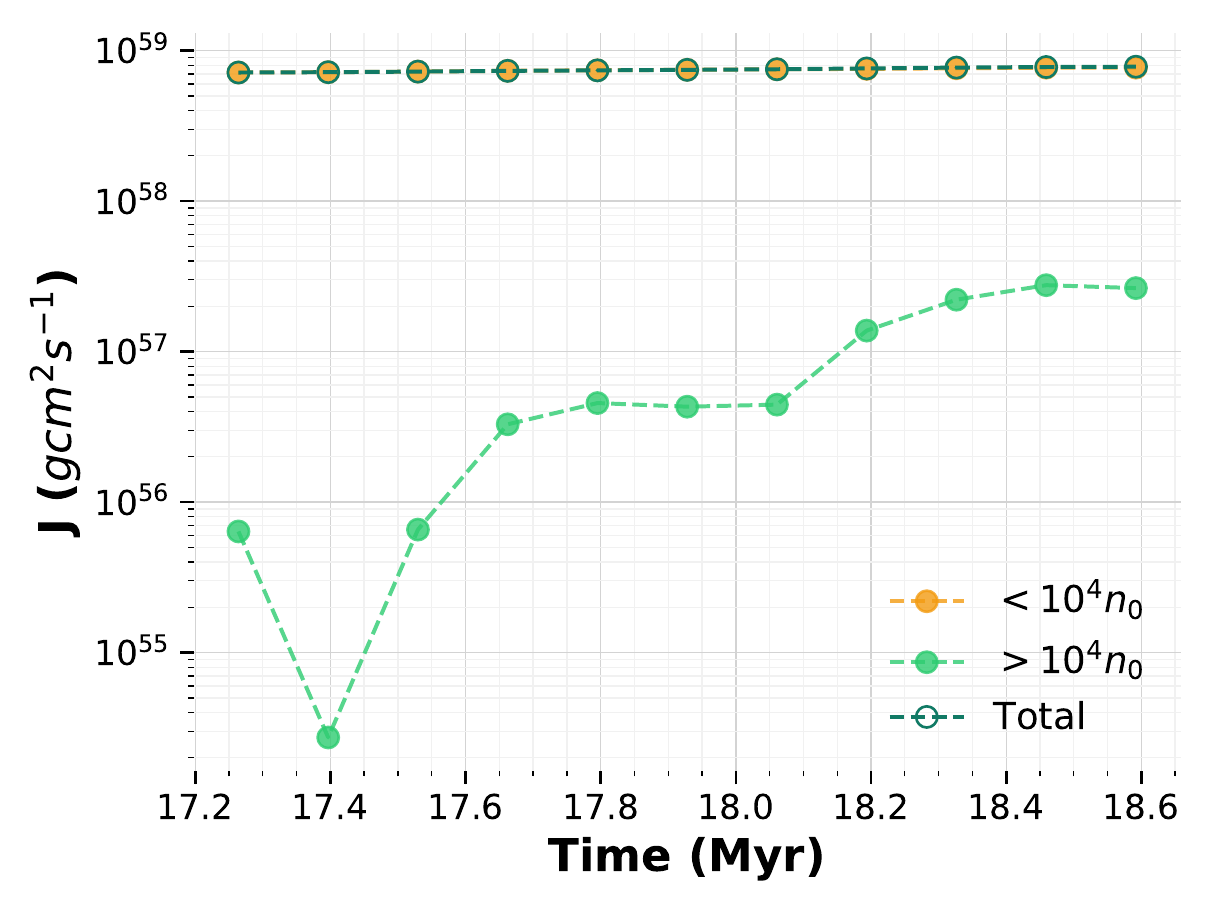}
  \includegraphics[width=0.3\textwidth]{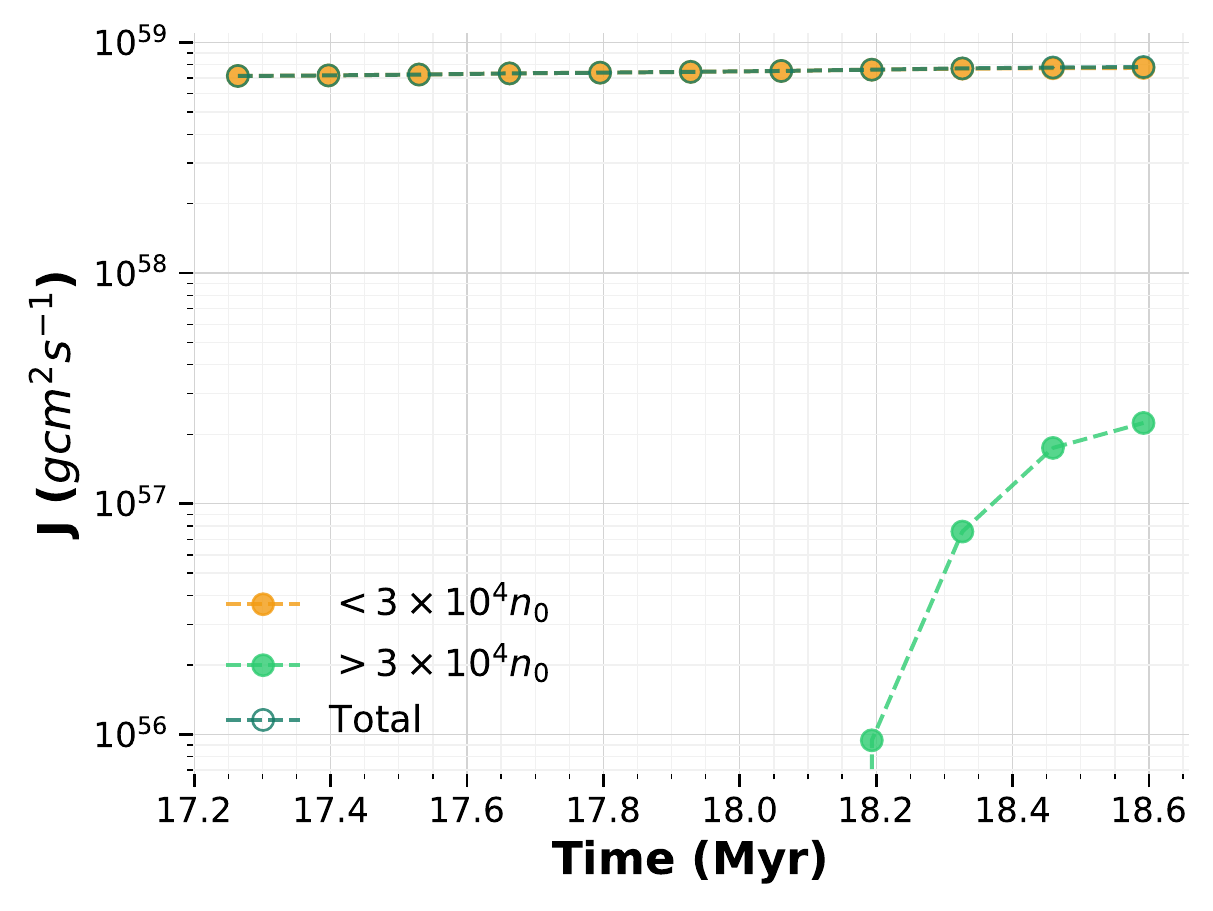}
  \hfill
  \includegraphics[width=0.3\textwidth]{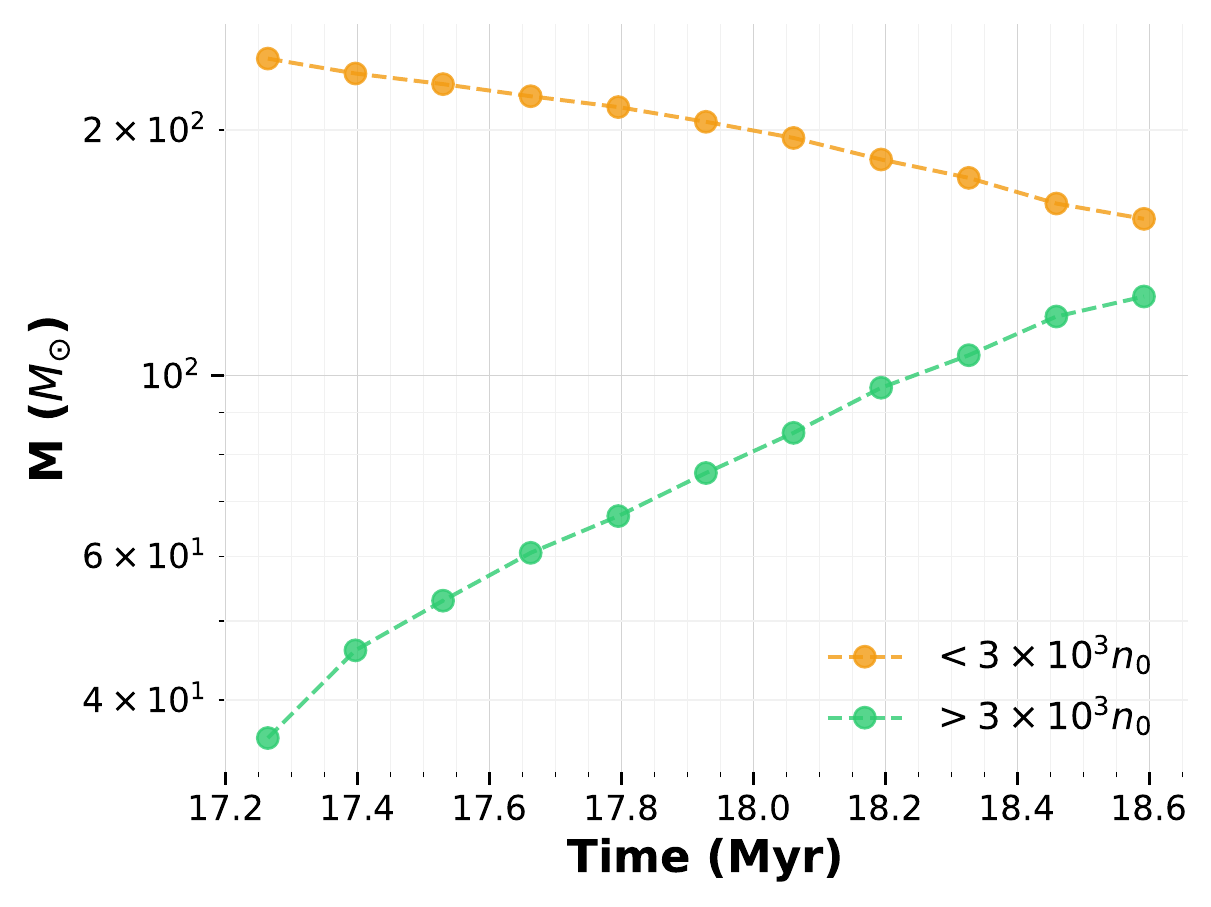}
  \includegraphics[width=0.3\textwidth]{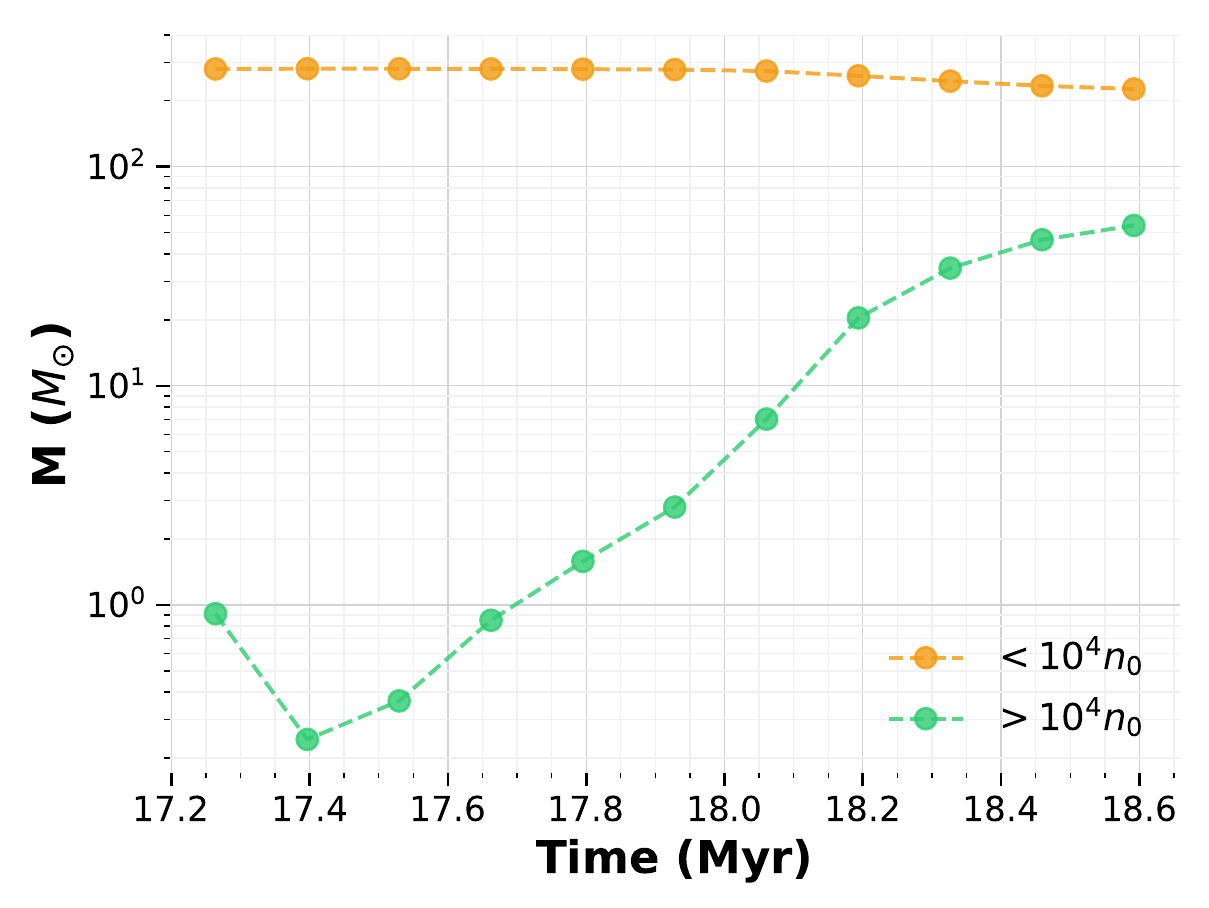}
  \includegraphics[width=0.3\textwidth]{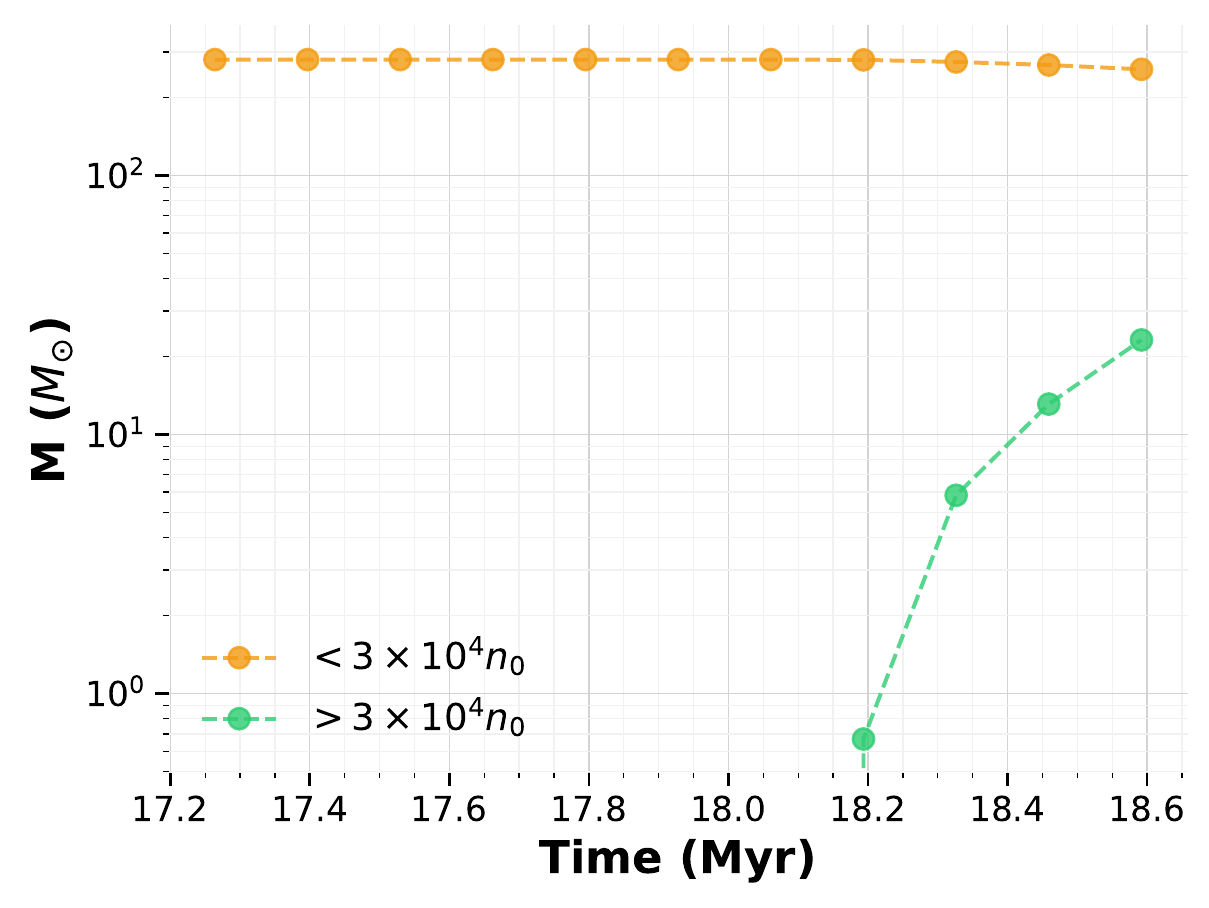}
  \caption{Evolution of the specific angular momentum $j$ (SAM, top row), the total angular momentum $J$ (AM, middle row) and the mass ($M$, bottom row), over a period of $1.33$ Myr, of the member particles of C11 (shown in Fig.\ \ref{fig:muestra one future}) that are above and below the density thresholds $n_{\rm th} = 3 \times 10^{3}$ (left column), $10^{4}$ (middle column) and $3 \times 10^{4}\, \pcc$ (right column). Note that, in the right column, there are no particles above the threshold before $t = 18.2$ Myr. For each threshold, the mass of the dense subset (light green) is seen to increase faster than its total AM, causing its SAM to decrease after some early transients. The opposite occurs to the low-density subset (orange), which undergoes a steady increase in its SAM during the second half of the time interval considered. The open dark green circles in the second row represent the total AM of the clump. The low-density region is thus seen to contain the vast majority of the total set's AM. The AM of each subset was calculated with respect to the center of mass of the full set.}
    \label{fig:separ}
\end{figure*}

\subsection{Interpreting the results} \label{sec:interp}

In the previous sections we have shown that the lagrangian sets of particles that make up a clump at some {\it final} time $\tdef$ tended to evolve close to the observational slope in the \jR\ diagram at relatively distant earlier times, when their member particles were significantly scattered, and the volume they occupied contained a high fraction of intruder particles. However, at times closer to $\tdef$, when they already constituted a nearly connected set, with relatively few intruder particles, the particle sets tended to evolve with $j \sim$ cst. On the other hand, true ``clumps'', defined as connected particle sets above a certain density threshold at all times during their evolution, evolve along the observational slope at all times. Finally, when tracking lagrangian sets towards the future, we found that not all of their particles participate in the collapse, and instead some of them (mostly the ones on the periphery) disperse away and decrease their density below the threshold density initially used to define the clump, effectively leaving the clump.

The above results suggest that a lagrangian set of SPH particles modifies its AM through the interaction with other neighboring particles, especially those interspersed among the member particles. Since the SPH particles essentially sample the flow at their locations, their masses are very small, and the density does not necessarily vary greatly among them, this form of AM exchange seems to correspond to hydrodynamic torques, rather than (or, at least, in addition) to the gravitational torques suggested by \citet{Larson84} and \citet{Jappsen+05}. These torques are exerted by shearing and compressive (in general, turbulent) stresses among the fluid parcels, as well as by the thermal pressure gradient (in general, the first and second terms on the right-hand side of eq.\ [\ref{eq:total torque}]). This mechanism of AM exchange through hydrodynamic torques is supported by the observation that only particle sets for which at some previous time $t < \tdef$ the number of intruders within the minimal rectangular box was larger than $3\times$ the number at $\tdef$ exhibit a period of evolution along the observational slope in the \jR\ diagram. 

This interpretation is also supported by the observation that the clump defined as a connected set above a density threshold throughout its evolution (Fig. \ref{fig:muestra same tresh}), so that it contains different sets of SPH particles at different times, evolves essentially parallel to the observational slope in the \jR\ diagram at all times (Fig. \ref{fig:evo same tresh}). Interestingly, however, since the defining density threshold is held constant, this clump actually increases its mass, size and AM (both total and specific) over time, due to accretion from its environment, in spite of harboring a local center of collapse.

But in addition, since not all of the clump's particles participate in the collapse, and instead a part of it is dispersed, it appears that the redistribution of AM occurs in a manner similar to that in accretion disks: if the whole object is subject to its self-gravity at all times, but prevented from contraction by the net rotation, then any local loss of AM in some subregion will allow it to contract, at the expense of transferring it to the rest of the parent structure. In this sense, the process of AM exchange is akin to fragmentation: the parcels that lose AM are the ones that contract. 

This in turn suggests that it is incorrect to think of a dense core as the result of the monolithic gravitational contraction of a larger clump, a process which should conserve AM. Instead, we suggest that the objects making up the observational data in Fig.\ \ref{fig:data} are precisely the {\it fragments} of larger structures subject to strong self-gravity that have managed to contract gravitationally because they have shed some of its AM via interactions with their neighboring fluid parcels. In this case, the observed \jR\ relation does not pose a ``problem'', but is just the natural result of turbulent torques being applied among fluid parcels, all subject to strong self-gravity, so that the parcels that lose AM are able to contract further. 

An important implication of this proposed mechanism is that a whole clump can never collapse in full. Instead, only a fraction of it can collapse, while the rest of its mass must be expelled, in order to carry the excess AM with it. This may impose an upper limit to the mass efficiency of fragmentation into denser units. In the case of accretion disks, it is well known that most of the mass is accreted onto the central object, while a vanishing amount of mass migrates outwards, carrying most of the AM. In the next section we measure the corresponding mass fractions in the case of the fragmentation of a clump.



\subsection{Testing the AM transfer mechanism} \label{sec:testing_mechanism}

In Sec.\ \ref{sec:interp} we have proposed that the observed \jR\ relation is the result of the exchange of AM among the fluid parcels making up a turbulent clump which is under the influence of its self-gravity, so that the portion of the clump that loses AM contracts, while the part that gains it expands. This mechanism is consistent with the GHC scenario \citep{VS+19}, in which molecular clouds are globally dominated by their self-gravity, although they still contain moderate turbulence that produces density fluctuations. In this scenario, the global mean Jeans mass decreases over time due to the global contraction \citep{Hoyle1953}, causing fluctuations of ever smaller masses to begin their own local collapse process as time proceeds. Here we suggest that the turbulence also causes AM transfer among the fluid parcels making up the clumps, and that conservation of the total AM limits the fraction of the cloud material that can continue to contract, since it must shed part of its mass to expel part of its AM. 
This mechanism is thus similar to that operating in accretion disks, but it operates in the amorphous, gaseous phase.

To support this suggestion, in Fig.\ \ref{fig:separ} we show the evolution of the AM of the dense (light green lines and dots) and diffuse (orange lines and dots) parts of the lagrangian particle set labeled C11 in Fig.\ \ref{fig:muestra one future}. At each temporal snapshot (indicated by the dots along the lines), we separately consider the particles with densities larger and smaller than the thresholds $3 \times 10^{3}$ (left column), $10^{4}$ (middle column) and $3 \times 10^{4}\, \pcc$  (third column), computing the SAM (top row), total AM (middle row), and mass (bottom row) of the high- and low-density subsets. The AM is computed for each group with respect to the center of mass of the entire clump. It can be seen that not only the low-density particles have values of $j$ several times higher than the dense ones, but in fact the low-density particle set tends to increase its SAM over time, while the high-density set tends to reduce it in general, except for a few transients. 
This reinforces the suggestion that, as the particles exchange AM through turbulent torques, the ones losing it can fall deeper into the gravitational potential well, becoming denser and more compact.

\subsection{A gravity-driven model for the constancy of $\beta$ and the \jR\ scaling}

\begin{figure}
\centering
\includegraphics[width=.48\textwidth]{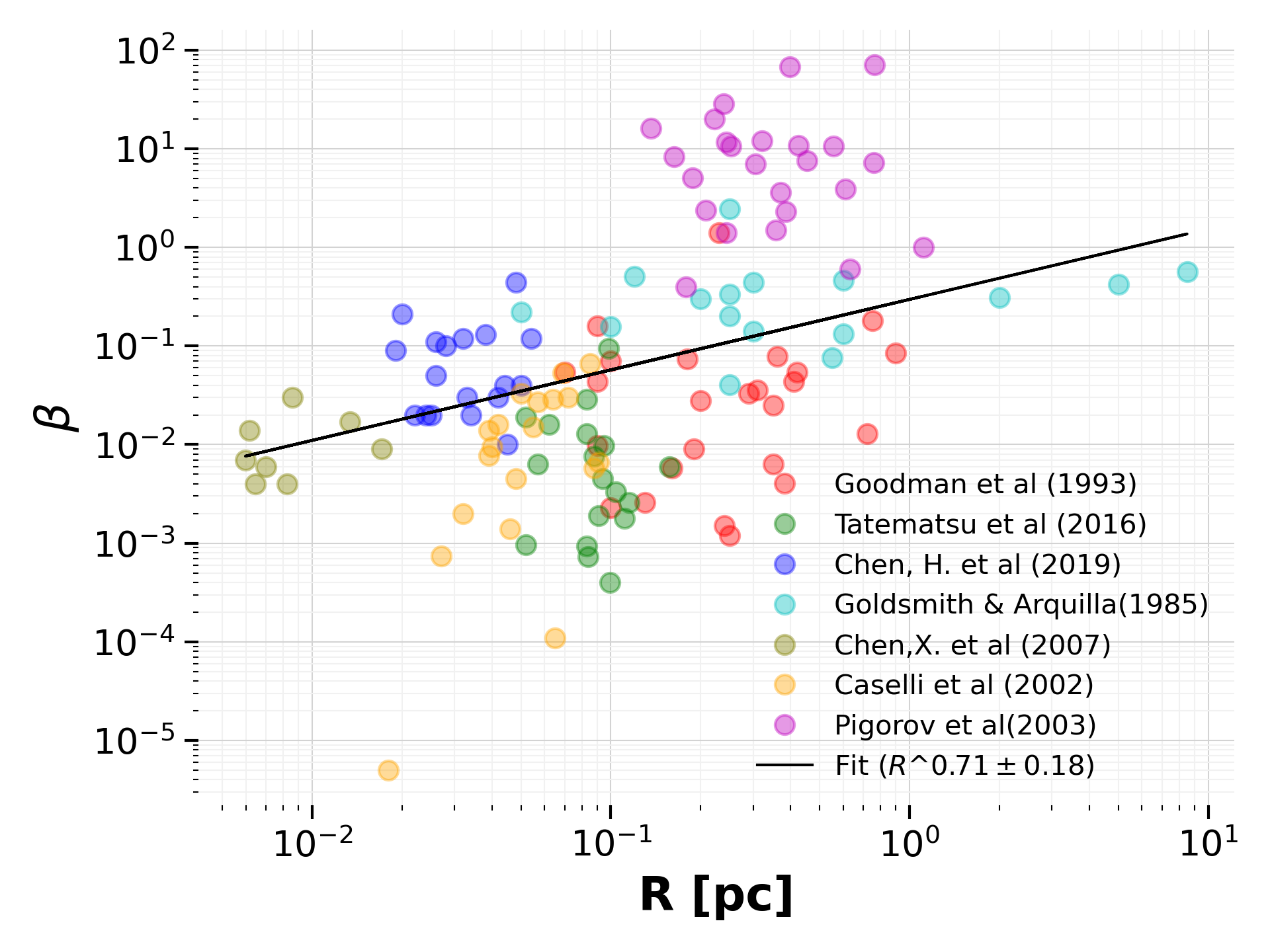}
\includegraphics[width=.48\textwidth]{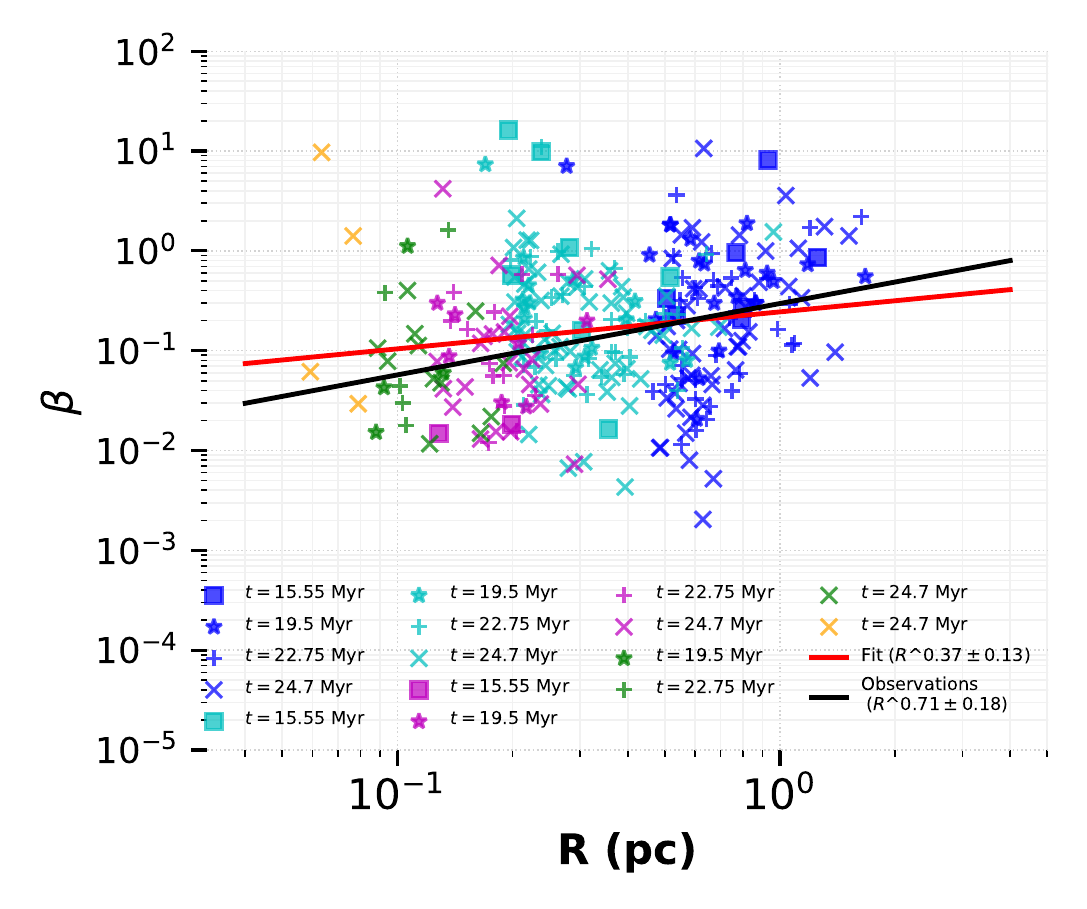}
 \caption{{\it Top panel:} $\beta$, the ratio of the rotational to the gravitational energy, {\it vs.}\ the radius $R$, for the observational sample. The black line represents a least squares fit to these data. {\it Bottom panel:} The same plot for the numerical sample. The black and red lines respectively represent the fits to the observational and the numerical data. Given the large scatter, the two samples appear to have very similar distributions.}
 \label{fig:numerical sample 2}
\end{figure}

In the previous sections we have presented evidence that the observed apparent loss of SAM is the result of the turbulent exchange of AM among the fluid parcels of a clump. However, the origin of the numerical value of the slope observed in the \jR\ plot is still not well understood. A few decades ago, \citet{Goodman+93} proposed a semi-analytical derivation of the dependence of the AM and the clump's radius based on assuming the \citet{Larson81} linewidth-size relation (which was thought to arise from virial equilibrium) and the empirical observation that $\beta$, the ratio of the rotational kinetic energy to the gravitational energy appears to be independent of the clump's radius. This property is approximately observed in both our compiled observational data and in our numerical sample, as shown in Fig.\ \ref{fig:numerical sample 2}. From these properties, they obtained a scaling relation of the form $j \propto R^{3/2}$, which was very close to the scaling found in their dense core sample (with exponent $\sim 1.6$) and to the slope fitted in this work for our compilation of observational data ($\sim 1.43$).

%

However, at present it is generally accepted that the Larson relations have been superseded by the \citet{Heyer+09} relation, of the form 
\begin{equation}
\sigma \propto (\Sigma R)^{1/2},
\label{eq:Heyer09}
\end{equation}
from which Larson's linewidth-size and density-size relations follow for samples of objects selected so that their column density is approximately constant \citep{BP+11, BP+12}. Additionally, within the context of the GHC scenario, assumed in this work, this scaling arises not from the virial equilibrium condition, $E_{\rm k} \approx |E_{\rm g}|/2$, but from the free fall condition, $E_{\rm k} \approx |E_{\rm g}|$, and applies not only to molecular clouds, but to the massive clumps and dense cores within the clouds \citep{BP+11, BP+18}. 


The self-similar, scale-free nature of the gravitational contraction process in the range of scales above the Jeans length may possibly explain the observed independence of $\beta$ with clump radius \citep{Goodman+93, Xu-Xuefang+2020b} as follows. Let us assume that all regions larger than the Jeans length are attempting to contract due to the domination of self-gravity, and consider a region of fixed mass $M$. Its gravitational energy per unit mass, $\eg$, then, scales with radius as
\beq
\eg \approx \frac{GM} {R} \propto R^{-1},
\label{eq:eg-R}
\eeq
while its specific rotational energy is
\beq
\er \approx \frac{1} {2} \frac{I \omega^2} {M} \approx \frac{1} {2} \vrot^2.
\label{eq:vrot}
\eeq
On the other hand, if the AM is conserved during the contraction,
\beq
J \approx {\rm cst.} \approx I \omega \approx M R^2 \frac{\vrot} {R} = M R \vrot,
\label{eq:cst_J}
\eeq
implying that $\vrot \propto R^{-1}$, and therefore
\beq
\er \propto R^{-2}.
\label{eq:erot-R}
\eeq

We thus see that, as a fluid parcel contracts due to gravity, the ratio of its rotational energy to its gravitational energy $\beta$ would tend to increase, if its AM were conserved. This could only continue until $\er \sim \eg$, at which point the collapse should be halted by rotation. However, if the parcel sheds its AM via turbulent torques, then the contraction can continue. That is, on the one hand, the combined action of AM conservation and gravitational contraction tends to increase the $\beta$ ratio, while on the other hand, the exchange of AM between the fluid parcel and its neighbours tends to counter this growth. It thus seems plausible that the competition between these two processes tends to keep $\beta$ approximately constant, or at least, independent of radius. In a future contribution, we plan to investigate the dependence of the rate of AM transfer on the gradient of the rotational energy density.

We can then perform a calculation similar to that of \citet{Goodman+93}, but without assuming Larson's velocity dispersion-size relation, and instead using the gravitational energy directly. We start by explicitly writing $\beta$ as the ratio of the rotational to gravitational energies, both per unit mass, denoted by $\er$ y $\eg$. Noting that $\er \approx (1/2) I \omega^2/M \approx (1/2) j\omega$ and $\eg \approx GM/R = \pi G R \Sigma $, where $\omega$ is a representative angular velocity for the clump and $\Sigma \approx M/\pi R^2$, we have 
\begin{equation}
\beta \equiv \frac{\er}{\eg} \approx \frac{1/2\, j \omega} {\pi G R \Sigma}.
\label{eq:beta_GHC}
\end{equation}
Under the assumption that $\beta \approx$ cst., we can then solve for $\omega$ as
\begin{equation}
\omega \approx \frac{2 \pi \beta G R \Sigma} {j}.
\label{eq:omega}
\end{equation}
On the other hand, the SAM is $j = I \omega / M \approx R^2 \omega$. Substituting eq. (\ref{eq:omega}) in this expression for $j$, we finally get
\begin{equation}
\label{eq:final}
j \approx (2 \pi \beta G \Sigma)^{1/2} R^{3/2},
\end{equation}
so that in the context of the GHC scenario and the \citet{Heyer+09} relation, we recover the dependence of $j$ on $R$, assuming that $\Sigma$ does not exhibit any specific trend with $R$, and that rotation draws its energy from the gravitational one---in agreement with the fundamental premise of the GHC scenario---and loses it by AM transfer.

\begin{figure*}
\centering
\includegraphics[width=\textwidth]{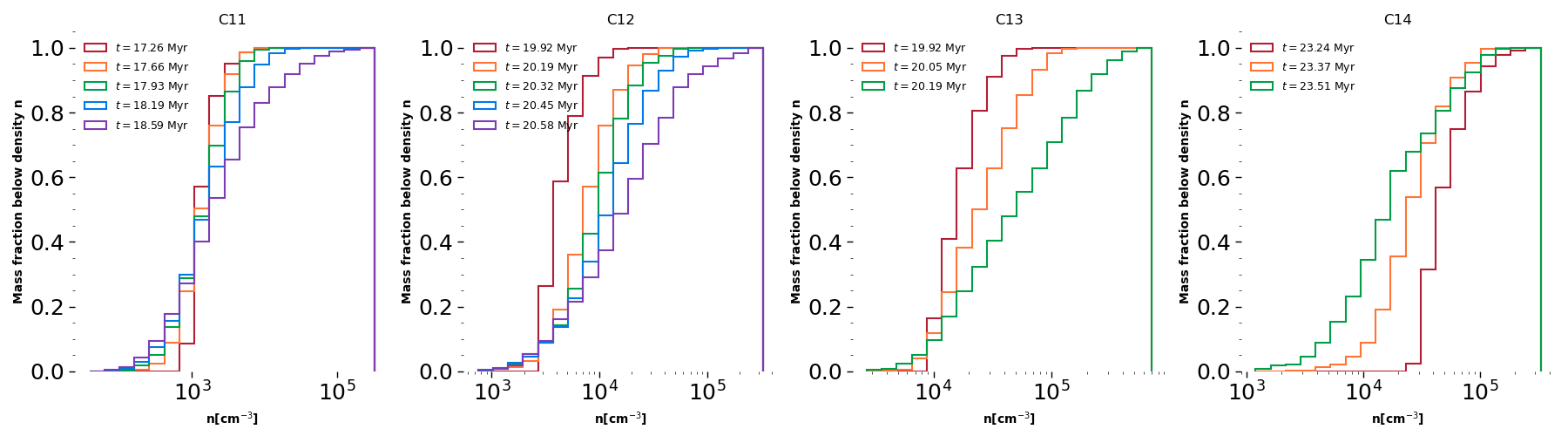}
 \caption{Evolution of the normalized cumulative density histogram for clumps C11 to C14 (shown in Fig.\ \ref{fig:muestra one future}), respectively defined with density thresholds $n_{\rm th} = 10^{3}\, \pcc$ (C11), $n_{\rm th} = 3 \times 10^{3}\, \pcc$ (C12), $n_{\rm th} = 10^{4}\, \pcc$ (C13), and $n_{\rm th} = 3 \times 10^{4}\, \pcc$ (C14), at times $\tdef = 17.26$ Myr for C11, $\tdef = 19.92$ Myr for C12 and C13, and $\tdef = 23.24$ for C14. It is seen that, at the final time (which represents the last snapshot prior to the formation of sinks) in the various cases, a fraction between $\sim 5$ and $60\%$ of the mass acquires a density below the definition threshold density, so that only the remaining mass can continue to collapse.}
 \label{fig:hist_dens}
\end{figure*}

\subsection{A limit on the mass available for collapse due to angular momentum conservation} 

In Secs.\ \ref{subsubsec:future} and \ref{sec:testing_mechanism}, when tracking the member particles of a clump to the future, we found that a fraction of them always leaves the clump, decreasing their density and dispersing to more distant locations. This mechanism takes AM away from the clump, and allows the remaining particles to become denser. But, in addition, this mechanism imposes an upper bound to the efficiency of the contraction mechanism, since not all of the clump's mass can reach a higher-density state. To quantify this upper bound, in Fig.\ \ref{fig:hist_dens} we show the evolution, from $\tdef$ to the last snapshot before the formation of a sink, of the mass fraction below the indicated density for the lagrangian particle sets corresponding to clumps C11 to C14 from Fig.\ \ref{fig:muestra one future}. These clumps were defined with a different density threshold each. 

It is seen from Fig.\ \ref{fig:hist_dens} that, for all four lagrangian sets, the mass fraction below the clump's defining threshold density increases monotonically in time, reaching $\sim 20\%$ of the total mass of the lagrangian set for C11 after 1.33 Myr, $\sim 5\%$ for C12 after 0.66 Myr, $\sim 5\%$ for C13 after $0.27$ Myr, and $\sim 60\%$ for C14, also after 0.27 Myr. Therefore, the mass loss needed to conserve AM may account for at least part of the observed core-formation efficiency of $\sim 30\%$ in star-forming regions \citep[e.g.,][]{Motte+98, Bontemps+2010, Palau+2013, Palau+2015}.


\section{Conclusions}
\label{sec:conclusions}

In this paper we have investigated the angular momentum exchange mechanisms among sets of fluid particles in an SPH simulation of dense cloud formation in the turbulent warm atomic ISM. Our strategy has profited from the particle nature of the SPH scheme, which allowed us to track over time the particle sets that constituted a ``clump'' (a connected set of particles above some threshold density $n_{\rm th}$) at some time $\tdef$. We referred to these as the ``member (or lagrangian) particle sets''. The tracking was performed either from the past ($t < \tdef$) or towards the future ($t > \tdef$), allowing us to see where the clump member particles came from, and how they evolve subsequently, and to measure their total (AM) and specific (SAM, denoted $j$) angular momentum. For comparison, we also tracked clumps defined in the traditional way (as connected particle sets above the threshold density) throughout their evolution. We also compared our results with the properties of a sample of observed clumps in various clouds, taken from the literature.

Our results can be summarized as follows:

\begin{itemize}

\item The numerical clump sample from our simulation of globally and hierarchically contracting clouds driven by self gravity reproduces with remarkable accuracy the observed \jR\ relation, showing that the GHC scenario produces clouds and clumps with a realistic AM content and scaling.

\item The loss of AM in molecular clouds and their substructures (referred to generically as ``clumps'') can be provided by the various torques acting on a fluid parcel of the cloud, described by the terms on the right-hand side of eq.\ \eqref{eq:total torque}, including the widely discussed magnetic, gravitational, and viscous torques. However, in an inherently turbulent medium such as molecular clouds, the torques due to the Reynolds stress and the pressure gradient terms in the momentum equation (first and second terms in eq.\ [\ref{eq:total torque}]), cannot be neglected as alternative important AM exchange channels. These may likely even be dominant in the interaction among fluid parcels of similar densities at small scales.


\item The lagrangian particle sets tracked from the past often exhibit two different possible evolutionary regimes: an early stage ($t$ significantly smaller than $\tdef$) of evolution along a trajectory roughly parallel to the locus of the observational sample of clumps in the \jR\ diagram, (thus losing SAM) and a later one ($t$ closer to $\tdef$) in which $j$ remains approximately constant.

\item The lagrangian sets tracked from the past that exhibited an AM loss stage were found to contain a large number (over three times the number at $\tdef$) of ``intruder'' particles within the minimal rectangular 3D box containing all the particles of the member set. Instead, the particle sets that evolved with $j \sim $ cst.\ throughout the tracking period contained a smaller (less than twice the number at $\tdef$) number of intruder particles in the minimal rectangular box since the start of the tracking period. This suggests that the evolution at $j \sim $ cst.\ occurs when the particle set does not have enough companion fluid particles to exchange the AM with. Also, this argues against the dominant torques being gravitational, since these should act over long distances, and thus should not require the companions to be nearby. 

\item In all the lagrangian sets tracked to the future, the SPH particles that reduced their density also moved away from the center of mass, effectively being lost from the dense clump, even if the latter contains a collapse center. Moreover, these particles contained most of the AM of the particle set. This mechanism is therefore qualitatively similar to that taking place in accretion disks, in which part of the mass moves outwards, carrying AM out, and allowing the rest of the mass to fall further inwards. In addition, the evolution in the $j$-$R$ diagram of these sets does not contain an abrupt change in slope as in the case of tracking from the past, lacking a constant-$j$ evolution period, thus suggesting that their dynamics is dominated by self-interactions between their constituent particles. 

\item Contrary to the case of accretion disks, we have found that the fraction of mass that is lost from the member particle sets  ranges from a few to over 50\%. It thus appears that the AM removal process in molecular clumps may be significantly more mass-consuming than the equivalent process in accretion disks, possibly contributing importantly to the low observed core formation efficiency. Further investigation is necessary to establish the relative importance of this mechanism in setting the efficiency.

\item On the other hand, clumps defined as connected sets above a density threshold throughout the tracking time seem to always evolve along the observed relation \jR, increasing both their radius and their mass while doing so. 
This is not surprising, as these clumps are equivalent to the parts of lagrangian sets that are able to contract by giving their AM to their neighbors.

\item At the time of their definition as connected particle sets above a threshold, the clumps in the simulation do not exhibit a significant dependence of $\beta$, the ratio of the rotational to the gravitational energy, with their size $R$, similarly to most observational samples.

\item We have suggested that the near independence of $\beta$ with the size $R$ of the clumps may be due to the tendency of the rotational energy to increase faster than the gravitational energy during gravitational contraction if $J \sim$ cst.\ (cf.\ eqs.\ [\ref{eq:erot-R}] and [\ref{eq:eg-R}]), causing a tendency of $\beta$ to increase. If the efficiency of AM exchange increases as the rotational energy density gradient in a clump increases, this may tend to bring $\beta$ to a stationary value, independently of size.


\end{itemize}

The above results are consistent with the evolution of all parcels within a dense cloud being driven by their self-gravity, but only those that can transfer some of their angular momentum to their neighboring fluid parcels are able to contract gravitationally, while those that receive it are expelled from the clumps, reducing their density, and failing to participate in the collapse. This process is therefore the proposed mechanism within the GHC scenario for molecular clouds, in which the whole cloud is dominated by its self-gravity, and thus prone to develop collapse at multiple scales within it.

This interpretation would then suggest that the so called ``angular momentum problem'' is nonexistent, because the collection of objects entering the observed \jR\ relation have not contracted monolithically to reach their present sizes, but rather have {\it fragmented} out of larger objects, and contracted precisely because they are the parts that have lost some of their AM to their neighbours. In this sense, the observation of the densest fragments in a cloud or clump constitutes a selection effect that picks up precisely the fragments which have lost AM.

\section*{Acknowledgements}

We thank Javier Ballesteros-Paredes for pointing out the similarity of the proposed AM transfer mechanism in the gaseous phase with that operating in accretion disks, and Gilberto G\'omez for useful discussions. This work has been supported in part by a CONACYT graduate fellowship for G.A.-C. 











\end{document}